\documentclass{ws-mpla}
\usepackage{graphicx}
\usepackage{caption}
\usepackage{xcolor}
\usepackage{longtable}
\usepackage[left]{lineno}

\usepackage[colorlinks=true, linkcolor=blue, citecolor=blue]{hyperref}
\usepackage[numbers,sort&compress]{natbib} 
\begin{document}

\markboth{S. Das et al.}{Rare Event Searches Using Cryogenic Detectors via Direct Detection Methods}

\catchline{}{}{}{}{}

\title{Rare Event Searches Using Cryogenic Detectors via Direct Detection Methods}

\author{S.	Das,
R.	Dey,
V. K. S.	Kashyap,
B.	Mohanty,
D.	Mondal}
\address{National Institute of Science Education and Research Bhubaneswar, An OCC of Homi Bhabha National Institute, Jatni 752050, Odisha, India \\bedanga@niser.ac.in}

\author{S.	Banik}
\address{Marietta-Blau-Institut f\"{u}r Teilchenphysik Institut der \"{O}sterreichischen Akademie der Wissenschaften, 1010 Wien, Austria and\\ Atominstitut, Technische Universit\"{a}t Wien, 1020 Wien, Austria}

\author{M.	Chaudhuri}
\address{Institute of Modern Physics, Fudan University, Shanghai 200433, China}

\author{V. Iyer}
\address{Department of Physics, University of Toronto, Toronto, ON M5S 1A7, Canada}

\maketitle


\begin{abstract}
 Cryogenic detectors are at the forefront of rare-event search experiments, including direct detection of dark matter, coherent elastic neutrino–nucleus scattering, neutrinoless double-beta decay, and searches for fractionally charged particles. Their unique ability to achieve ultra-low energy thresholds, typically $\mathcal{O}$(eV–100 eV), together with excellent energy resolution and effective background suppression, makes them sensitive to extremely faint signals from rare interactions. These rare particle interactions produce phonons, ionization, or scintillation, depending on the target medium, which are registered by specialized sensors and converted into measurable signals. This review summarizes the underlying detection principles, surveys major experiments and recent results, examines forthcoming initiatives, and outlines the evolving role of cryogenic detectors in advancing the frontiers of rare-event physics.

\end{abstract}


\section{Introduction}	

The search for rare events represents one of the most compelling frontiers in modern experimental physics, aiming to directly detect phenomena that probe both the Standard Model (SM) and physics beyond it. Examples include dark matter (DM) interactions, coherent elastic neutrino–nucleus scattering (CE$\nu$NS), search for fractionally charged particles (FCP) and neutrinoless double-beta decay ($0\nu\beta\beta$). These processes are characterized by extremely small interaction probabilities, with cross-sections typically below $10^{-35}$ cm$^2$. Unlike astrophysical observations that infer new physics indirectly, such as galaxy rotation curve anomalies \cite{first_proposed_DM:1933gu}, gravitational lensing \cite{Clowe2006_bullet_cluster}, and cosmic microwave background observations \cite{planck_res_2018}, direct detection seeks unambiguous signatures of particle interactions in highly sensitive detectors operated under well-controlled laboratory conditions.

Because the expected event rates and associated energy depositions are exceedingly small, typically $\mathcal{O}$(keV) or below, suppressing background noise is paramount. Operating detectors at cryogenic temperatures (often below 100 K) substantially reduces thermal fluctuations (since thermal fluctuation scales linearly with temperature) and electronic noise. This enables a faithful measurement of weak signals from such rare events. This capability has made cryogenic technologies central to many leading rare-event search programs, from DM and CE$\nu$NS to $0\nu\beta\beta$ decay.

This review focuses on the role of cryogenic detectors in direct detection searches for rare-event candidates. We begin by outlining the detection principles of phonon-, ionization-, and scintillation-based cryogenic systems (Section~\ref{sec:cryogenic_detector_technology}). We then survey experimental efforts, starting with DM, covering weakly interacting massive particles (WIMPs) as well as lighter candidates such as axion-like particles (ALPs) and dark photons (Section~\ref{sec:dark_matter}), followed by searches for fractionally charged particles (FCPs; Section~\ref{sec:FCPs}), CE$\nu$NS (Section~\ref{sec:CEvNS}), and $0\nu\beta\beta$ (Section~\ref{sec:NDBD}). Finally, Section~\ref{sec:summary} summarizes the current status and outlines future directions in the field.

\section{Cryogenic Detectors in Search for Rare Events\label{sec:cryogenic_detector_technology}}
In the field of rare-event searches, cryogenic detectors have become popular because they combine exceptional energy resolution with ultra-low thresholds, enabling sensitivity to signals well beyond the reach of conventional technologies~\cite{intro,cryo}. This makes them uniquely suited to detecting the minute energy depositions arising from processes such as DM interactions, CE$\nu$NS scattering, and $0\nu\beta\beta$ decay.

\begin{figure}[h]
    \centering
\includegraphics[width=0.6\linewidth]{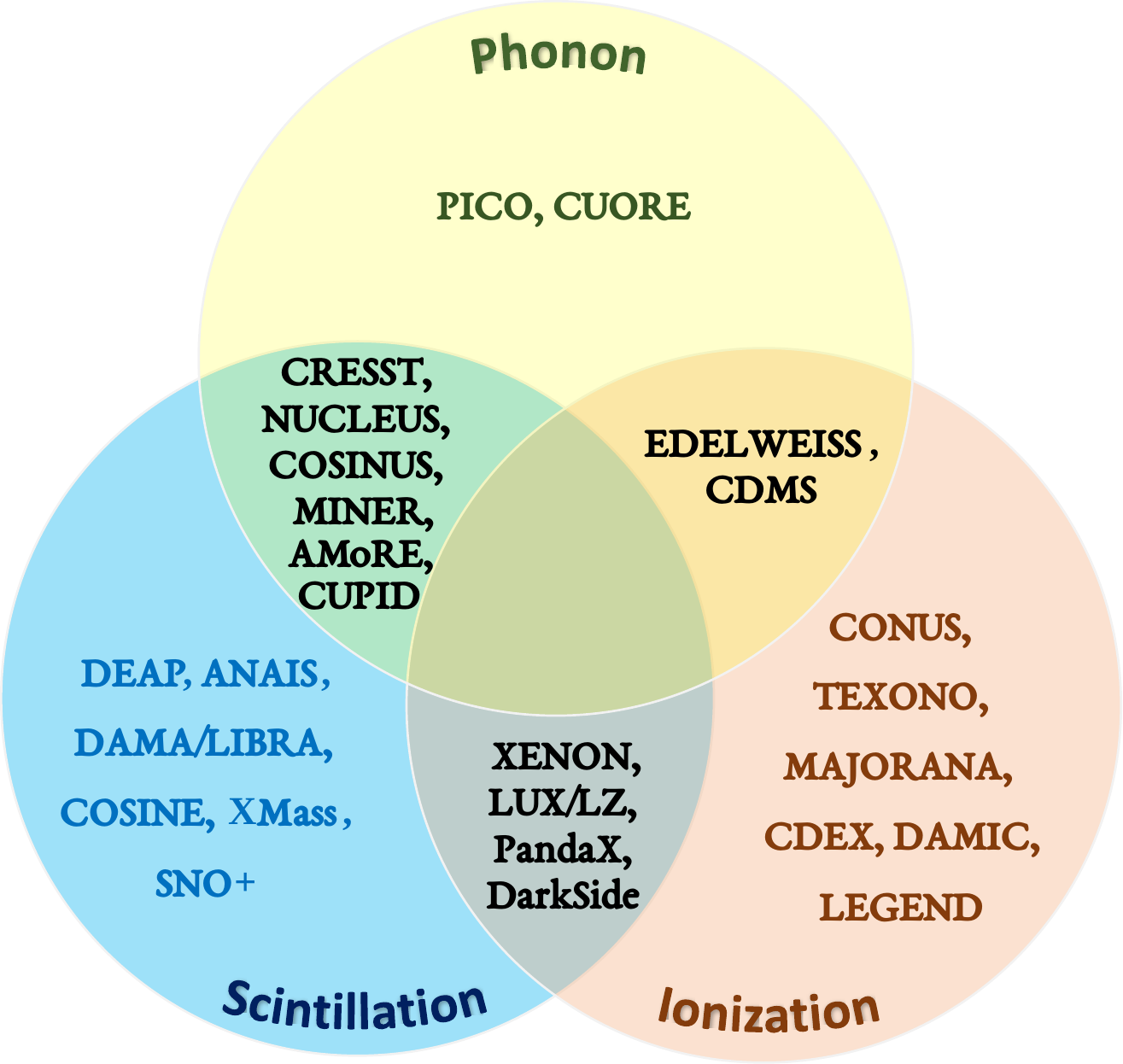}
    \caption{\label{fig1} Venn diagram of direct detection experiments, based on the signal detection: phonons or heat (yellow), scintillation or light (blue), and ionization or charge (orange).}
    \end{figure}

In 1984, Fiorini and Niinikoski~\cite{fiorini1984} proposed the use of cryogenic calorimeters for high-sensitivity experiments in elementary particle physics, including $0\nu\beta\beta$ decay and neutrino mass measurements. The first experimental realization of this concept in rare-event searches came with the Cryogenic Rare Event Search with Superconducting Thermometers (CRESST) experiment, which employed sapphire ($\mathrm{Al_{2}O_{3}}$) calorimeters to detect phonon signals from particle interactions~\cite{history_cresst}. Soon after, the Cryogenic Dark Matter Search (CDMS) experiment advanced the technique by developing germanium (Ge) and silicon (Si) detectors capable of simultaneously measuring ionization and phonons, greatly improving the discrimination between nuclear and electronic recoils~\cite{history_cdms,cresst_cawo4}. To enhance background rejection and DM sensitivity, CRESST later adopted calcium tungstate ($\mathrm{CaWO_4}$) targets~\cite{cresst_cawo4}, which yield both scintillation light and phonon signals upon particle interaction. Not all phonon-based detectors employ dual readout, for example, CUORE operates purely as a calorimeter relying solely on phonon measurements~\cite{dr_cresst}.

Cryogenic temperatures of $\sim$100 K are typically achieved using liquefied gases such as nitrogen (N$_2$), argon (Ar), or xenon (Xe). In these systems, the detector may either be thermally coupled to the cryogen, or the cryogen itself can act as the active detection medium. For even greater sensitivity to low-energy recoils using phonon-based detection, detectors are further cooled down to millikelvin (mK) temperatures using dilution refrigerators, which exploit the enthalpy of mixing between $\mathrm{^3He}$ and $\mathrm{^4He}$ isotopes~\cite{dr}. Below $\sim$0.87 K, the mixture separates into a $\mathrm{^3He}$-rich concentrated phase and a $\mathrm{^3He}$-poor dilute phase~\cite{dr_cresst}. Continuous cooling is achieved as $\mathrm{^3He}$ atoms transition from the concentrated to the dilute phase, absorbing heat from the surroundings.

At these temperatures, thermal noise is drastically suppressed, enabling the detection of energy depositions as small as a few electronvolts (eVs), well below the thresholds of traditional detectors~\cite{tinkham_superconductivity}. However, dilution refrigerator-based cryogenic technologies face scalability challenges when compared to other direct detection methods such as sodium iodide (NaI) crystals or noble-liquid detectors (LAr, LXe), which more readily achieve tonne-scale exposures. Typical detectors measure low-energy depositions via three signal channels: ionization (charge), scintillation (light), and phonons (heat). Examples of experiments utilizing these detection modes are illustrated in Fig.~\ref{fig1}.

While all three approaches exploit the advantages of cryogenic operation, each offers distinct strengths, limitations, and design considerations. To date, no detector has simultaneously measured all three signal channels, which would have provided a stronger handle on background rejection. The following subsections describe the working principles of phonon, ionization, and scintillation based cryogenic detectors, highlighting their respective roles in rare-event searches. 

\subsection{Phonon-based cryogenic detectors\label{phonon_detector}} 
When a particle interacts with a cryogenic detector material, it deposits energy that generates a population of primary phonons along with electron-hole ($e^{-}$–$h^{+}$) pairs. If an electric field ($E$) is applied in the detector, these charge carriers can drift through the lattice, producing secondary phonons, known as Neganov–Trofimov–Luke (NTL) phonons~\cite{LUKE2,LUKE1} or Luke phonons, through lattice interactions, thereby amplifying the total phonon signal. The resulting phonons are collected by specialized sensors, photo-lithographically patterned on the detector surface.

Several types of phonon sensors have been developed, each with distinct advantages and limitations. The most widely used are Neutron Transmutation Doped (NTD) thermistors and Transition Edge Sensors (TES)~\cite{fabjan}.

\begin{figure} [h!]
\centering
  \includegraphics[width=0.9\linewidth]{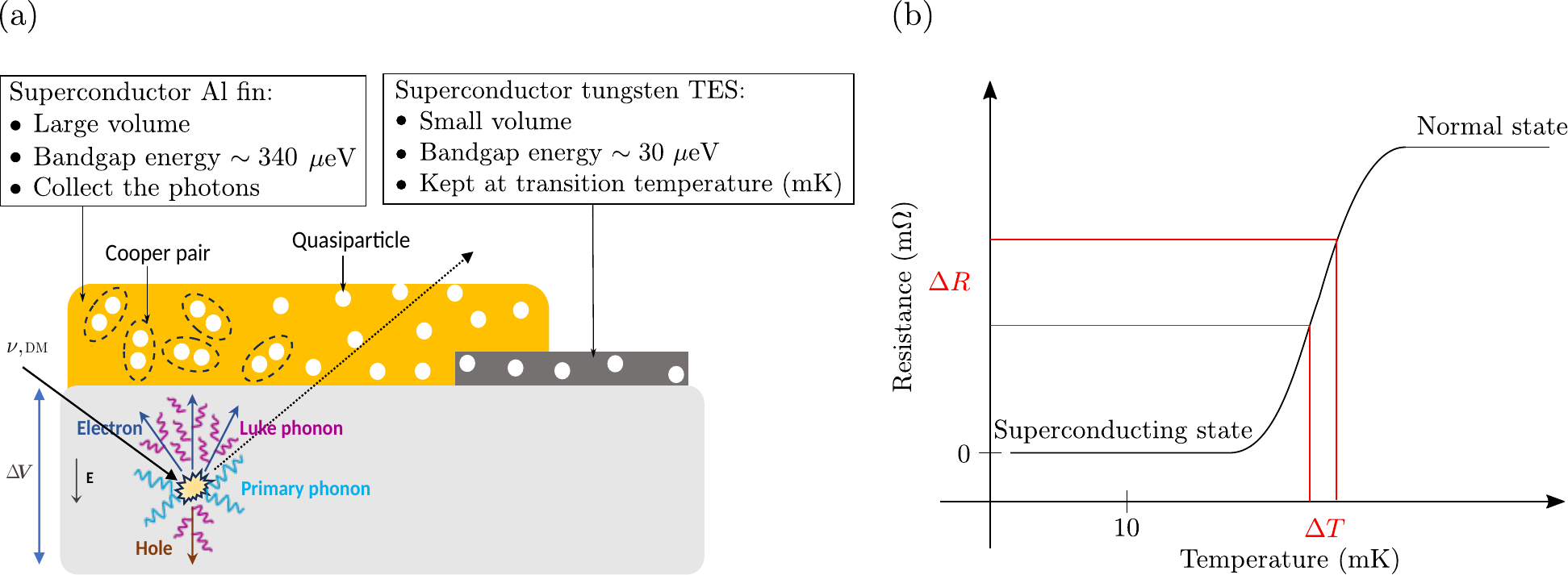}
  \caption{\label{fig2}(a) A Schematic diagram of a phonon-based cryogenic detector with TES sensor. (b) Typical TES transition: Resistance vs Temperature near $\mathrm{T_{C}}$.}
  \label{fig:detector_QET_schematic}
\end{figure}

\begin{figure}
\centering
  \includegraphics[width=\linewidth]{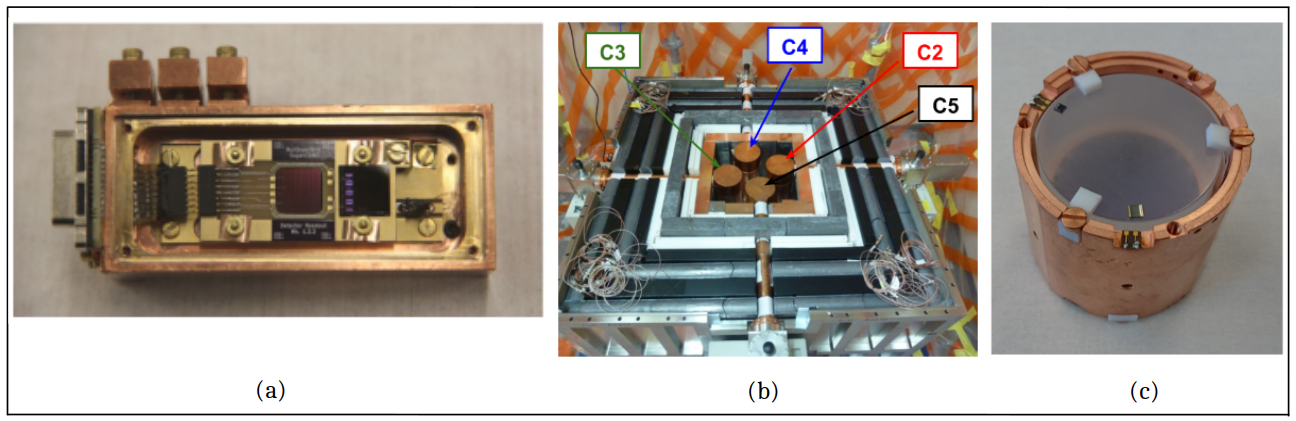}
  \caption{\label{fig3} The High Voltage eV-scale detector used by  SuperCDMS~\cite{hvev,SuperCDMS_SNOALB_SNOWMASS_2021}. (b) A p-type point-contact (PPC) Ge detector, labels C1–C5 are explained in the reference~\cite{ppc}. (c) A cryogenic detector at CUPID that measures scintillation and phonon signals~\cite{cupid_det}.}
  \label{fig:Cryo_Detectors}
\end{figure}

\begin{itemize} 
    \item \textbf{Neutron Transmutation Doped (NTD) Ge Thermistors:}
    Among the earliest phonon sensors used in cryogenic detectors, including early CDMS prototypes, NTD thermistors operate by exploiting the sharp temperature dependence of electrical resistance in doped semiconductors, typically Ge, around 10-100 mK temperatures. Doping is precisely achieved via neutron irradiation in a nuclear reactor, producing a uniform dopant distribution, which makes its resistivity strongly temperature dependent. For example, in the CUORE experiment, each $\mathrm{TeO_2}$ crystal is instrumented with a NTD-Ge thermistor glued on its surface, allowing it to measure the minute temperature rise of the absorber and convert it into an electrical signal~\cite{cuore1,cuore2,cuore3}. When a particle interacts in the absorber crystal, the deposited energy quickly thermalizes into phonons, producing a small temperature increase ($\Delta T$). Due to the steep $R(T)$ dependence, this results in a large measurable resistance change in the NTD sensor. The resistance variation is read out with a low-noise circuit, typically using a superconducting quantum interference device (SQUID) amplifier for high sensitivity~\cite{SQUID}. NTD thermistors are stable, relatively easy and inexpensive to fabricate, and well suited for large-mass bolometers such as the CUORE and CUPID detectors~\cite{cuore_ntd,cupid_ntd}. However, they offer slower response times and somewhat poorer energy resolution compared to more advanced technologies.

    \item \textbf{Transition Edge Sensors (TES):}
    Extensively deployed in experiments such as SuperCDMS and CRESST~\cite{tes_cresst}, TES devices are superconducting tungsten (W) films operated near their critical temperature ($T_\mathrm{C}$), the fixed temperature at which a material transitions from superconducting to normal condition, typically $\sim$10 mK. Operating in this narrow region near $T_\mathrm{C}$ allows even tiny energy depositions to produce measurable resistance changes. Large area aluminum (Al) fins having a higher $T_\mathrm{C}$ of about 1.2 K, are coupled to the TES to collect incoming phonons. Phonon absorption in the Al fins breaks Cooper pairs, creating quasiparticles, [Fig.~\ref{fig2}(a)]. The minimum phonon energy required for quasiparticle creation in Al is $\sim$ 340 $\mu$eV, corresponding to twice the superconducting gap, $\mathrm{E_{min}}$ = 1.76 $\mathrm{k_{B}T_{C}}$, here $\mathrm{k_{B}}$ is the Boltzmann constant, as given in the reference~\cite{tes,tinkham_superconductivity}. These quasiparticles diffuse into the W TES, where the band gap energy is much smaller ($\sim$30~$\mu$eV), allowing efficient energy deposition. Some loss occurs at interfaces, due to acoustic impedance mismatch between the absorber and sensor, and due to quasiparticle trapping or recombination at the Al–W boundary, reducing overall signal transmission efficiency. Despite these losses, even a minute energy input raises the TES temperature enough to induce a sharp change in resistance near $T_\mathrm{C}$ [Fig.~\ref{fig2}(b)]. This resistance change alters the current ($\mathcal{O}(\mu\mathrm{A})$) through a SQUID readout circuit, enabling precision phonon measurement. To achieve high sensitivity, TES sensors must be maintained in their narrow transition region. A TES-based cryogenic High Voltage eV-scale (HVeV) detector is shown [Fig.~\ref{fig3}(a)].
\end{itemize}

While both NTD thermistors and TES devices are mature and reliable phonon-sensing technologies, they occupy distinct niches in rare-event searches. NTDs excel in large-mass bolometric applications where stability and scalability are critical, as in CUORE, whereas TES sensors offer faster response times, superior energy resolution, and lower thresholds, making them the preferred choice for low-mass DM searches and experiments requiring event-by-event particle discrimination. This complementarity has driven the parallel evolution of phonon-sensing technologies alongside ionization and scintillation-based cryogenic detection methods, discussed in the following subsections.

Metallic Magnetic Calorimeters (MMCs), as employed by the AMoRE experiment, represent another powerful cryogenic sensor technology for rare event searches~\cite{amore}. In addition to NTD and TES technologies, emerging superconducting sensors such as Microwave Kinetic Inductance Detectors (MKIDs)~\cite{mkid} and Superconducting Nanowire Single Photon Detectors (SNSPDs) (including related microwire variants)~\cite{snspd} are being developed for phonon and low-energy detection.

\subsection{Ionization-based cryogenic detectors\label{sec:ionization_based}}
Ionization-based cryogenic detectors operate by collecting electron–hole ($e^{-}$–$h^{+}$) or electron–ion pairs generated by particle interactions within the target material. Typical detector volume include high-purity semiconductor crystals such as Si and Ge, as well as noble elements like Xe and Ar, operated at low temperatures ($\sim$ 100 K or below) to suppress thermal noise. Detector performance depends critically on ionization yield, intrinsic radiopurity, energy resolution, and scalability for rare-event searches.

Among these technologies, high-purity germanium (HPGe) detectors stand out due to their excellent charge transport properties (longer carrier lifetimes and higher mobility), combined with low intrinsic radioactivity and outstanding energy resolution~\cite{ppc_2}. These attributes make HPGe detectors particularly effective for probing low-energy signatures from GeV-scale DM, CE$\nu$NS, and $0\nu\beta\beta$ decay. This is further aided by the favorable $Q$-value (the total energy released in the decay) of ${}^{76}$Ge at 2039 keV~\cite{gerda}, which lies above most naturally occurring $\gamma$-ray lines and thereby reduces the background from environmental radioactivity.

A major advancement in cryogenic ionization detection is the development of the p-type point-contact (PPC) HPGe detector [Fig.~\ref{fig3}(b)]. The PPC detectors achieve sub-keV energy thresholds, ultra-low net impurity concentrations ($\sim$10$^{8}$ atoms/cm$^{3}$), and excellent pulse-shape discrimination. They have been deployed in leading rare-event searches, including CONUS~\cite{conus}, TEXONO~\cite{texono}, MAJORANA~\cite{majorana}, and CDEX~\cite{cdex}. The PPC configuration consists of a p-type Ge crystal with a boron-implanted p$^{+}$ point contact at the center (signal readout electrode) and a lithium-diffused n$^{+}$ contact covering most of the outer surface (high-voltage electrode). A positive bias applied to the n$^{+}$ contact, with the point contact grounded, causes electrons to drift toward the n$^{+}$ electrode while holes are collected at the p$^{+}$ contact~\cite{ppc_1}.

Ionizing interactions within the depletion region produce $e^{-}$–$h^{+}$ pairs in number proportion to the deposited energy. At 77 K, an average of 2.96 eV energy is required to produce a single $e^{-}$–$h^{+}$ pair in Ge crystal, which exceeds the band gap ($\sim$0.67 eV) because part of the deposited energy is dissipated through non-ionizing processes, primarily phonon excitation~\cite{Pehl,Wei}. This corresponds to a mean yield of $\sim$340 pairs/keV for electron recoils. For nuclear recoils, such as those produced in CE$\nu$NS or DM scattering, only a fraction of the recoil energy contributes to ionization. This suppression, described by the quenching factor (typically $\sim$0.2 for Ge), depends on both the recoil energy and material properties~\cite{Bonhomme,Scholz}.

The small point contact in PPC detectors results in a capacitance of only 1–2 pF at full depletion~\cite{ppc_2,Bonet}, which significantly reduces electronic noise and leads to improved energy resolution. The ultra-low impurity concentration leads to slightly non-uniform electric fields across the detector, causing charge carriers to drift at different velocities depending on their location. This produces position-dependent pulse shapes, allowing discrimination between single-site events, where all energy is deposited locally, typical of $0\nu\beta\beta$ signals and multi-site events, where energy is deposited in multiple locations, mostly from background $\gamma$-rays (e.g., $\mathrm{Tl^{208}}$)~\cite{ppc_1}. By analyzing the pulse shape at the point contact, PPC detectors can effectively separate signal-like single-site events from background multi-site events. Signal formation follows the Shockley–Ramo theorem~\cite{Shockley_1,Shockley_2}, which describes the induced charge $Q(t)$ at the electrode due to the motion of charge carriers,
\begin{equation}\label{eq:WP}
Q(t) = q_{0} \mathrm{[WP(\overrightarrow{r_{h}}(t)) - WP(\overrightarrow{r_{e}}(t))]},
\end{equation}
where $\mathrm{WP(\overrightarrow{r}(t))}$ is the weighting potential at the instantaneous position of the drifting holes or electrons, $\overrightarrow{r_h}(t)$ and $\overrightarrow{r_e}(t)$, respectively~\cite{Bonet}. Once both carriers are collected, the total induced charge reaches $Q(t) = q_0$, determined primarily by the holes. A charge-sensitive preamplifier at the point contact integrates the induced current, yielding a voltage signal proportional to the deposited energy.

In summary, ionization-based cryogenic detectors, particularly PPC HPGe devices, offer exceptional energy resolution, low energy thresholds, and powerful background discrimination, making them indispensable for precision measurements in rare-event searches. While phonon-based detectors excel at capturing the total deposited energy with ultra-low thresholds, ionization-based systems provide complementary information on the charge yield, enabling powerful event-type classification when combined with phonon or scintillation readouts. This complementary combination underlies many hybrid detector designs, and forms a natural bridge to the scintillation-based cryogenic detection techniques discussed in the following subsection.

\subsection{Scintillation-based detectors}
Scintillation-based detectors are a key technology in rare-event searches, including CE$\nu$NS and DM interactions. In these systems, the target material emits photons when traversed by energetic particles. Energy deposition excites electrons to higher energy states, subsequent de-excitation leads to the emission of scintillation photons, typically in the visible or near-UV range. The scintillation yield, the number of photons produced per unit deposited energy, depends on the particle type and the target medium~\cite{knoll}. Nuclear recoils generally produce significantly less light than electron recoils due to the quenching effect, where a larger fraction of the recoil energy is lost to phonon excitation~\cite{Lindhard,AprileDoke}. This difference in the light-to-heat ratio enables event-by-event particle discrimination, a critical capability for rare-event searches.

In dilution refrigerator-based cryogenic operation, scintillating crystals such as calcium tungstate ($\mathrm{CaWO_4}$) and sapphire ($\mathrm{Al_2O_3}$) are widely used. For instance, the MINER experiment~\cite{MINER_sapphire_scintillation} employs $\mathrm{Al_2O_3}$ to probe reactor-based CE$\nu$NS, while CRESST~\cite{cresst} and NUCLEUS~\cite{NUCLEUS_bkg} utilize $\mathrm{CaWO_4}$ and $\mathrm{Al_2O_3}$ detectors that simultaneously measure both phonons and scintillation light. In addition, several $0\nu\beta\beta$ decay experiments employ scintillating crystals such as TeO$_2$, Li$_2$MoO$_4$, CaMoO$_4$, and ZnSe as target materials, taking advantage of their dual phonon-light readout technique, as used in the CUPID experiment [Fig.~\ref{fig3}(c)], for background discrimination. These setups operate at temperatures below 20 mK, typically with TES sensors optimized for low-energy phonon detection. In particular, the CRESST-III detector module consists of a 24~g $\mathrm{CaWO_4}$ crystal ($20 \times 20 \times 10$ mm$^3$) serving as the phonon detector (PD), coupled to a silicon-on-sapphire disc ($20 \times 20 \times 0.4$ mm$^3$) acting as the light detector (LD)~\cite{LD}. When particles interact in the sapphire detector, they produce scintillation photons with a wavelength of around 325 nm~\cite{LD1}. When these photons are absorbed in the Si detector, they generate phonons and electron–hole pairs, which are subsequently measured by the TES. This two-channel readout enables active background discrimination by comparing the phonon and scintillation signals on an event-by-event basis.

By contrast, large-scale experiments such as XENONnT~\cite{xenonnt_2023} and PandaX-4T~\cite{pandax_2021} employ dual-phase liquid xenon time projection chambers (LXe-TPCs) at cryogenic temperatures. In these detectors, interactions in LXe produce prompt scintillation light (S1) and ionization electrons. The ionization electrons drift upward under the influence of an applied electric field (typically a few hundred V/cm), before being extracted into the gas phase, where they generate a secondary electroluminescence signal (S2). The ratio S2/S1 provides powerful discrimination between nuclear and electronic recoils. Other scintillation-based experiments operate at or near room temperature and rely solely on scintillation light for energy reconstruction. Notable examples are DAMA/LIBRA~\cite{dama} and COSINE-100~\cite{cosine}, both of which employ large arrays of NaI(Tl) crystals to investigate possible DM-induced annual modulation signals.

Cryogenic detection has matured into a highly diverse landscape, with each readout modality offering distinct strengths and weaknesses. TES-based phonon sensors and NTD bolometers have proven exceptionally successful in reaching sub-100~eV thresholds and enabling background rejection, yet the trade-off remains between response time, resolution, and scalability. Ionization-based PPC detectors achieve outstanding energy resolution but face persistent challenges from quenching-factor uncertainties and near-threshold calibration. Scintillation readout adds powerful particle identification but is limited by optical collection efficiency and impurity control. A concise comparison of the three primary cryogenic detection modalities, phonon, ionization, and scintillation based, is presented in Table~\ref{tab:cryogenic_modalities}. This summary highlights the distinct signal channels, target materials, operational strengths, and limitations of each detection modality. This multi-modal approach complements purely phonon or ionization-based systems, and sets the stage for the physics results discussed in the subsequent sections.

Looking ahead, hybrid systems combining multiple channels, as well as emerging superconducting sensors such as MKIDs and SNSPDs, are expected to push thresholds into the single-eV regime. Future experiments must prioritise precision calibration, scalable cryogenics, and systematic control if they are to maintain sensitivity gains in larger-mass detectors. 

\begin{table}[h]
\centering
\caption{Comparison of cryogenic detection modalities for rare-event searches.}
\label{tab:cryogenic_modalities}
\renewcommand{\arraystretch}{1.15}
\setlength{\tabcolsep}{3pt}
\resizebox{\linewidth}{!}{
\begin{tabular}{p{2.1cm} p{2.7cm} p{2.6cm} p{3.0cm} p{3.0cm}}
\hline
\textbf{Detection Modality} & \textbf{Primary Signal} & \textbf{Typical Targets} & \textbf{Advantages} & \textbf{Limitations} \\
\hline
\textbf{Phonon} & Athermal phonon & Ge, Si, $\mathrm{CaWO_4}$, $\mathrm{Al_2O_3}$. & Ultra-low thresholds (few eV); excellent $E$ resolution; largely independent of ionization yield unless operated under applied bias. & Weak intrinsic particle identification; Challenging to scale to tonne-scale detectors. \\
\textbf{Ionization} & Charge carriers ($\mathrm{e^{-}-h^{+}}$/$\mathrm{e^{-}}$-ion) & HPGe, Si, LXe, LAr. & Excellent $E$ resolution (HPGe); High charge collection efficiency; Event topology discrimination. & Higher $E$ thresholds vs. phonon.\\
\textbf{Scintillation} & Visible/UV photons & $\mathrm{CaWO_4}$, $\mathrm{Al_2O_3}$, NaI(Tl), LXe, LAr, $\mathrm{TeO_2}$, $\mathrm{Li_2MoO_4}$. & Nuclear/electron recoil separation (S2/S1); large-mass scalability. & Yield sensitive to impurities/quenching; photon collection limited by optical efficiency. \\
\hline
\end{tabular}
}
\end{table}

\section{Search for Dark Matter}\label{sec:dark_matter}

Dark matter (DM)~\cite{first_proposed_DM:1933gu,DM_review_paper_2024} is a hypothesized, non-luminous, non-baryonic component of the Universe that interacts primarily through gravity and possibly via additional interactions beyond the Standard Model (BSM). Its existence is strongly supported by multiple independent astrophysical and cosmological observations, including galaxy rotation curves~\cite{rotation_curve_1970}, gravitational lensing in the bullet cluster~\cite{bullet_cluster_2006}, and cosmic microwave background (CMB) measurements by the \textit{Planck} satellite~\cite{Planck_2015_CMB}. Collectively, these observations indicate that DM constitutes about five times more mass than ordinary baryonic matter and is composed of particles that are gravitationally interacting, electrically neutral, cold (i.e., non-relativistic), nearly collisionless, and stable over cosmological timescales.

Despite compelling evidence for its existence, the particle nature of DM remains unknown. Over the past few decades, it has become one of the most intensively studied subjects in rare-event searches. Experimental strategies to detect DM are broadly categorized into three approaches: indirect detection~\cite{indirect_search}, collider-based searches~\cite{collider_search}, and direct detection. In this review, we focus on direct detection with cryogenic detectors and assess their performance in comparison with other leading experimental techniques.

\begin{figure}[tbh]
\centering
\includegraphics[width=1\linewidth]{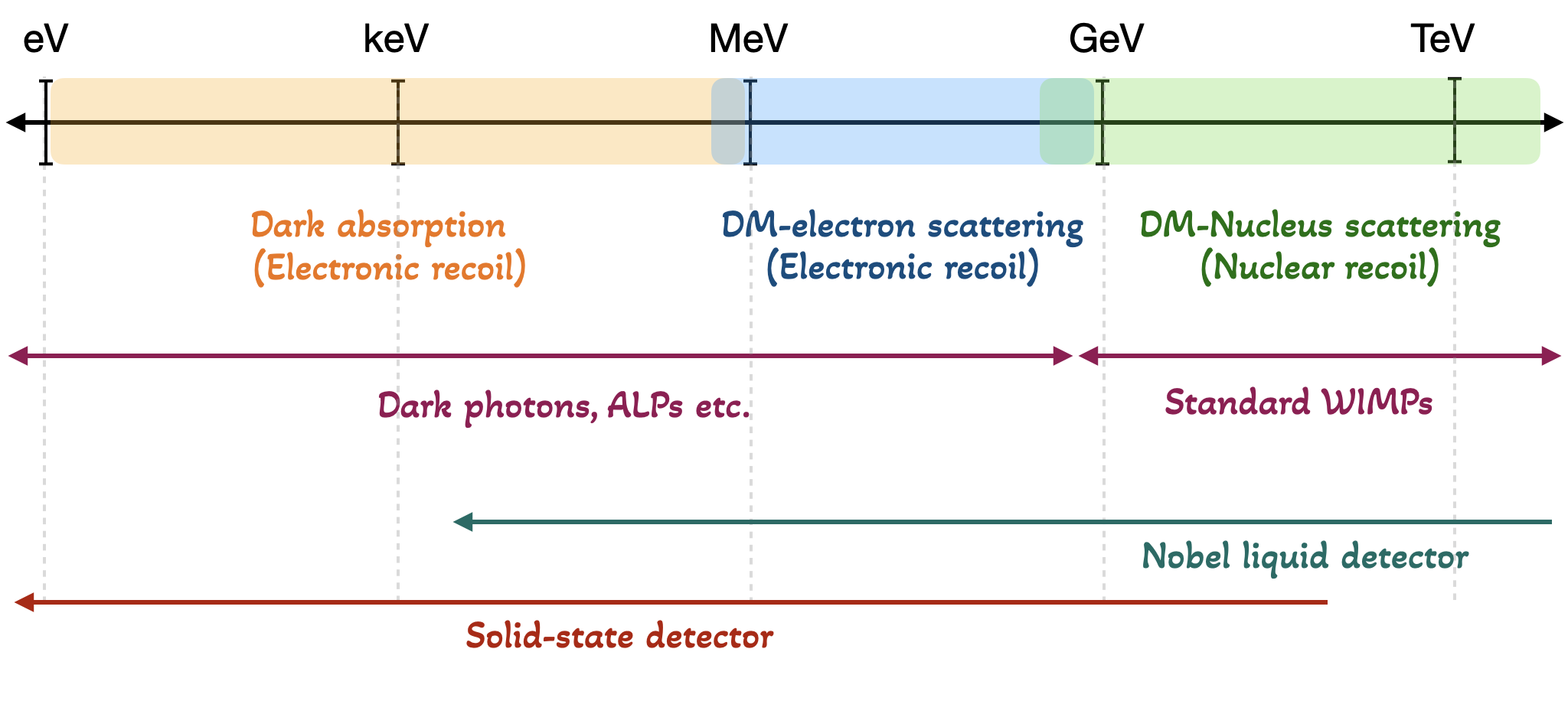}
\caption{Theoretically motivated dark matter candidates discussed in this article, along with their expected mass ranges and corresponding detection strategies.}
\label{fig:DM candidates}
\end{figure}

Because the mass of DM candidates is unconstrained, direct detection experiments explore a wide range of possible scales. In this article, we highlight three theoretically motivated candidates: (a) Weakly Interacting Massive Particles, predicted by many extensions of the SM; (b) axions--originally proposed to resolve the strong charge-parity (CP) problem and ALPs; and (c) dark photons, hypothetical gauge bosons that mediate interactions within a hidden or dark sector~\cite{Particle_DM_review_2005,DM_review_paper_2024}. Figure~\ref{fig:DM candidates} summarizes their expected mass ranges, possible interaction channels, and the corresponding detection strategies.

Table~\ref{WIMP_search_table} summarizes the leading direct detection experiments discussed in this review, highlighting their detector materials, measured signal type, and typical detection thresholds.

\begin{table}[h]
\begin{minipage}{\linewidth}
\centering
\caption{Major direct-detection dark matter experiments and their detection technologies. Experiments that primarily measure phonon signals achieve superior energy resolution compared to those relying mainly on scintillation or ionization. Simultaneous measurement of two channels enhances background discrimination. For energy thresholds, the subscripts \textit{nr} and \textit{ee} denote nuclear recoil energy and its electron-equivalent energy, respectively. The electron-equivalent energy is reduced relative to the nuclear recoil energy by a material-dependent quenching factor~\cite{lindhard_factor,lindhard_factor_deviation,birks_law_2013theory}.}
\label{WIMP_search_table}
\resizebox{\linewidth}{!}{%
\begin{tabular}{p{2.7 cm}p{3.5 cm}p{5 cm}p{3.2 cm} c}
\hline
\textbf{Experiment} & \textbf{Detector Material} & \textbf{Signal measured} & \textbf{Typical detection Threshold} & \textbf{Reference} \\ \hline
CDMS / SuperCDMS & Ge, Si & Phonon, Ionization & $\sim$70\,eV$_\mathrm{ee}$ & \cite{CDMSliteR2_WIMP,CDMSliteR3_WIMP_PLR_2019,SuperCDMS_migdal_Brem_WIMP_2023,SuperCDMS_SNOLAB_projected_sensitivity_2016,SCDMS_soudan_ALPs_DP,SuperCDMS_HVeV_R2_ALPs_DP,SuperCDMS_HVeV_R3_ALPs_DP} \\
EDELWEISS & Ge & Phonon, Ionization & $\sim$100\,eV$_\mathrm{ee}$ & \cite{EDELWEISS_WIMP_2019,EDELWEISS_migdal_2022,EDELWEISS_III_2018_ALPs_DP}\\
Majorana & Ge & Phonon, Ionization & $\sim$1\,keV$_\mathrm{ee}$ & \cite{Majorana_2017_ALPs}\\
CRESST & CaWO$_4$, Al$_2$O$_3$ & Phonon, Scintillation & $\sim$30\,eV$_\mathrm{nr}$ & \cite{CRESST_2019_WIMP,CRESST_2024_WIMP}\\
DarkSide & Liquid Argon (LAr) & Scintillation, Ionization (TPC) & $\sim$0.6--1.8\,keV$_\mathrm{nr}$ & \cite{DarkSide_2023_QF_WIMP_with_migdal,DarkSide_50_2023_ALPs_DP}\\
XENON & Liquid Xenon (LXe) & Scintillation, Ionization (TPC) & $\sim$1.0--2.0\,keV$_\mathrm{nr}$ & \cite{XENONnT_2025_WIMP,XENON_1T_2019_ALPs_DP} \\
LZ & Liquid Xenon (LXe) & Scintillation, Ionization (TPC) & $\sim$1.0--1.5\,keV$_\mathrm{nr}$ & \cite{LZ_2023_WIMP}\\
PandaX & Liquid Xenon (LXe) & Scintillation, Ionization (TPC) & $\sim$1.5--2.0\,keV$_\mathrm{nr}$ & \cite{PandaX_2025_WIMP,PandaXII_2017_ALPs}\\
PICO & Superheated C$_3$F$_8$ & Bubble Nucleation (bubble chamber) & $\sim$2--3\,keV$_\mathrm{nr}$ & \cite{PICO_WIMP_2019}\\
XMASS & Liquid Xenon (LXe) & Scintillation & $\sim$1.0\,keV$_\mathrm{ee}$ & \cite{XMASS_2018_ALPs_DP}\\
SENSEI & Silicon CCDs & Ionization & $\sim$1--2\,e$^-$ & \cite{SENSEI_2019_DP,SENSEI_2020_DP}\\
\hline
\end{tabular}%
}

\end{minipage}
\end{table}

\subsection{Search for Weakly Interacting Massive Particles (WIMPs)\label{sec:WIMP}}

The most widely studied candidate for dark matter is the WIMP, typically expected to have a mass in the range $\mathcal{O}(\mathrm{GeV/c^2})$ to $\mathcal{O}(\mathrm{TeV}/c^2)$~\cite{WIMP_1_1984,WIMP_2_1995_supersymmetric}. WIMPs could have been produced in the early Universe via the thermal freeze-out mechanism~\cite{Relic_WIMP_abundances_2012}. Theoretically motivated models with sub-GeV WIMPs also exist and are commonly referred to as low-mass or light dark matter~\cite{DM_review_paper_2024}. WIMPs can scatter elastically with detector nuclei, producing recoil energy ($E_R$) that can be measured experimentally. Direct detection experiments aim to constrain both spin-independent (SI) and spin-dependent (SD) WIMP--nucleon cross sections. SI interactions are governed by scalar mass couplings, while SD interactions arise from axial-vector spin couplings~\cite{SI_and_SD_WIMP_search}. Sensitivity to SD interactions requires target nuclei with non-zero net nuclear spin. Cryogenic targets such as Al$_2$O$_3$ and LiAlO$_2$, employed in the CRESST experiment, are well-suited for probing both SI and SD couplings. In contrast, silicon (Si) and germanium (Ge) detectors, such as those used in SuperCDMS and EDELWEISS, are primarily sensitive to SI interactions since their most abundant isotopes have zero nuclear spin. For consistency and comparability across different experiments, this review focuses exclusively on SI WIMP-nucleon cross-section limits.

The differential event rate for SI WIMP--nucleon scattering can be written as~\cite{Standard_halo_model}
\begin{equation}
\label{eq:SI_diff_rate}
    \frac{dR}{dE_R} =
    \frac{\rho_0 \, \sigma_n^{\mathrm{SI}}}{2 m_\chi \mu_N^2}
    \left( \frac{\mu_N}{\mu_n} \right)^2
    A^2 F^2(E_R) \, \eta(v_{\min}),
\end{equation}
where $\rho_0$ is the local dark matter density, $m_\chi$ is the WIMP mass, $\mu_n$ and $\mu_N$ are the reduced masses of the WIMP--nucleon and WIMP--nucleus systems, respectively, and $\sigma_n^{\mathrm{SI}}$ is the spin-independent WIMP--nucleon cross section. The parameter $A$ is the atomic mass number of the target nucleus, and $F(E_R)$ is the nuclear form factor, which accounts for the loss of coherence at finite momentum transfer. $\sigma_n^{\mathrm{SI}}$ is conventionally quoted at zero momentum transfer ($q = 0$, $F = 1$), enabling direct comparison across experiments. The function $\eta(v_{\min})$ denotes the mean inverse speed of WIMPs,
\begin{equation}
    \eta(v_{\min}) = \int_{v_{\min}}^{v_{\max}} \frac{f(\mathbf{v})}{v} \, d^3v,
\end{equation}
where $f(\mathbf{v})$ is the WIMP velocity distribution in the Earth's frame, normalized to unity. In the standard halo model, the DM halo is approximated as an isothermal, isotropic sphere with density decreasing as $r^{-2}$ from the Galactic center~\cite{Standard_halo_model}. Under this model, the local DM velocity distribution follows a truncated Maxwell--Boltzmann form in the Galactic frame. The minimum velocity required to generate a recoil energy $E_R$ is $v_{\min} = \sqrt{m_N E_R /(2 \mu_N^2)}$, where $m_N$ is the target nucleus mass. The upper cutoff, $v_{\max} \approx 600$ km/s~\cite{Standard_halo_model}, corresponds to the Galactic escape velocity. In addition to elastic scattering, modern direct detection experiments also explore inelastic WIMP-nucleon interactions, such as those arising from the Migdal effect and Bremsstrahlung emission~\cite{SuperCDMS_migdal_Brem_WIMP_2023}, which produce detectable secondary signals even when the nuclear recoil energy is below threshold, thereby extending sensitivity to sub-GeV WIMPs.

Direct detection experiments measure the nuclear recoil spectrum expected from WIMP interactions. Equation~\ref{eq:SI_diff_rate} provides the mapping between the measured event rate and the corresponding cross-section. In the absence of a statistically significant excess above background, 90\% confidence level upper limits on the WIMP--nucleon cross section is set, presented as exclusion curves in the $m_\chi$--$\sigma$ plane. No conclusive evidence for WIMPs has yet been found. Cryogenic detectors are particularly competitive and often world-leading in the low-mass regime ($\mathcal{O}$(GeV/$c^2$) and below), owing to their excellent energy resolution and $\mathcal{O}$(10 eV) thresholds. 

Systematic uncertainties in low-mass WIMP searches with cryogenic detectors arise primarily from the nuclear recoil quenching factor--ionization yield in phonon-ionization detectors or light yield in phonon-scintillation detectors--which governs both recoil energy reconstruction and event-type discrimination~\cite{Strauss:2014_light_quenching_CaWO4,SuperCDMS:ionisation_yield_CDMSlite_2022,SuperCDMS:ionisation_yield_Si_100eV_2023}. Additional contributions come from uncertainties in low-energy calibration (at $\mathcal{O}$(keV) and below) of phonon and secondary readout channels, underscoring the importance of precise calibration for sub-GeV dark matter searches. Recent studies~\cite{Ren:2020_NFC1_device_paper,SuperCDMS:2025_Compton_step} also show that detector response is strongly dependent on the applied biasing conditions: at high bias voltages, Neganov–Trofimov–Luke (NTL) phonons dominate the signal, whereas at zero bias the response is governed by primary phonons. Other sources of systematic uncertainty include signal modeling, background modeling, energy resolution, and event selection efficiencies \cite{SuperCDMS:sys_unc_quenching}. While the relative magnitudes of these effects vary across experiments, the quenching factor often provides the dominant contribution, particularly in sub-GeV searches. Overall, systematic uncertainties could be up to $\mathcal{O}$(10\%) level and are propagated into statistical analyses, generally leading to conservative or weaker exclusion limits \cite{SuperCDMS:sys_unc_quenching,CRESST_2019_WIMP}.

\begin{figure}[h!]
    \centering
    \includegraphics[width=0.9\linewidth]{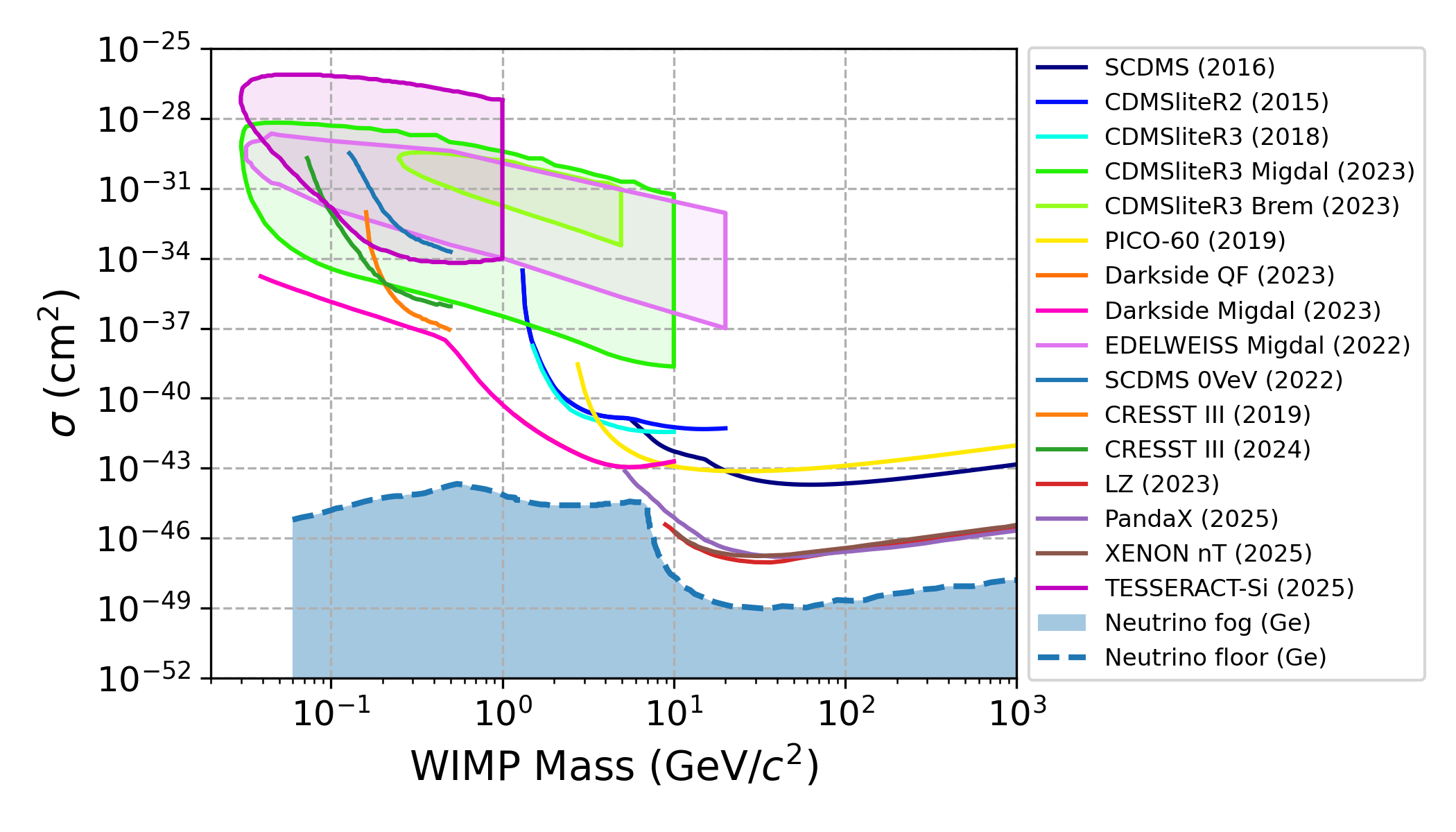}
    \caption{90\% confidence level upper limits on the WIMP--nucleon scattering cross section from SuperCDMS~\cite{CDMSliteR2_WIMP,CDMSliteR3_WIMP_PLR_2019,SuperCDMS_migdal_Brem_WIMP_2023}, EDELWEISS~\cite{EDELWEISS_WIMP_2019,EDELWEISS_migdal_2022},
    TESSERACT~\cite{TESSERACT:2025tfw}, PICO~\cite{PICO_WIMP_2019}, DarkSide~\cite{DarkSide_2023_QF_WIMP_with_migdal}, CRESST~\cite{CRESST_2019_WIMP,CRESST_2024_WIMP}, LZ~\cite{LZ_2023_WIMP}, PandaX~\cite{PandaX_2025_WIMP}, and XENON~\cite{XENONnT_2025_WIMP}, together with the neutrino floor for germanium targets~\cite{Billard:2021_APPEC_report_neutrino_floor_Ge}. Below 0.06~GeV/$c^{2}$, published calculations of the neutrino floor are unavailable, as they would require unrealistically low thresholds ($\mathcal{O}$(meV) or less) for Ge detectors and rely on uncertain low-energy neutrino flux predictions. Experiments such as CRESST and SuperCDMS dominate the low-mass regime, while dual-phase TPC experiments such as XENON, LZ, and PandaX set the most stringent limits at higher masses. Experimental limits are taken from public data or digitized from published figures.}
    \label{fig:WIMP_limit}
\end{figure}

Figure~\ref{fig:WIMP_limit} shows exclusion curves from leading direct detection experiments. CDMS~\cite{CDMSliteR2_WIMP,CDMSliteR3_WIMP_PLR_2019,SuperCDMS_migdal_Brem_WIMP_2023,SuperCDMS_SNOLAB_projected_sensitivity_2016}, EDELWEISS~\cite{EDELWEISS_WIMP_2019,EDELWEISS_migdal_2022}, TESSERACT~\cite{TESSERACT:2025tfw} and CRESST~\cite{CRESST_2019_WIMP,CRESST_2024_WIMP} primarily employ cryogenic phonon-based detectors. In contrast, DarkSide~\cite{DarkSide_2023_QF_WIMP_with_migdal} uses liquid argon TPCs, while XENON~\cite{XENONnT_2025_WIMP}, LZ~\cite{LZ_2023_WIMP}, and PandaX~\cite{PandaX_2025_WIMP} deploy dual-phase liquid xenon TPCs. PICO~\cite{PICO_WIMP_2019} utilizes superheated liquid bubble chambers. The ceiling observed in the closed exclusion curve arises from the limiting effect of the underground overburden. Limits from both elastic and inelastic scattering channels are shown. The most stringent limits from cryogenic detectors in the sub-GeV regime currently come from CRESST and TESSERACT~\cite{TESSERACT:2025tfw} (elastic-scattering–based searches) and SuperCDMS (inelastic-scattering–based searches). The blue dashed line in Fig.~\ref{fig:WIMP_limit} represents the so-called neutrino floor, the irreducible CE$\nu$NS background~\cite{Billard:2021_APPEC_report_neutrino_floor_Ge} beyond which a DM signal becomes indistinguishable. This boundary is for elastic scattering based WIMP search. Furthermore, recent studies~\cite{Neutrino_fog,neutrino_floor_reactor_antineutrino} have demonstrated that the neutrino floor is not a hard exclusion limit but rather a sensitivity threshold, below which neutrino backgrounds obscure potential DM signals, a regime often referred to as the neutrino fog. The exact location of this floor can also depend on the experimental site. In addition, the neutrino floor for inelastic WIMP searches based on the Migdal effect has recently been explored as well~\cite{migdal_neutrino_floor}.

Looking ahead, upcoming cryogenic experiments such as SuperCDMS SNOLAB~\cite{SuperCDMS_SNOLAB_projected_sensitivity_2016} and CRESST-III Phase~2~\cite{CRESST_2025_projected_sensitivity} are expected to significantly improve sensitivity, especially in the sub-GeV mass range. These advances will probe previously inaccessible parameter space in the search for low-mass dark matter candidates. 

Cryogenic experiments have decisively shaped the low-mass WIMP landscape, where their ultra-low thresholds outperform noble-liquid and bubble-chamber technologies. However, progress is increasingly limited by fundamental bottlenecks: uncertainties in quenching factors, sub-keV calibration, and irreducible CE$\nu$NS backgrounds near the neutrino fog. Inelastic signatures such as the Migdal effect enable sensitivity to even lower WIMP masses by exploiting ionization signals from inelastic scattering events that occur below the nuclear-recoil threshold. In addition, spectral and directional observables could help distinguish dark-matter interactions from neutrino scattering via the CE$\nu$NS proces.

\subsection{Search for Axion-Like Particles (ALPs) and Dark Photons\label{sec:ALPs}}

The lack of conclusive evidence for WIMPs has motivated growing interest in lighter dark matter candidates below the MeV/$c^2$ scale, such as axions, ALPs, and dark photons. Axions were originally proposed in quantum chromodynamics (QCD) to explain the absence of charge-parity CP violation in strong interactions~\cite{QCD_Axion}. Astrophysical constraints restrict the QCD axion mass to $\lesssim \mathcal{O}(10^{-2})$ eV/$c^2$~\cite{QCD_Axion_mass}, which lies below the typical detection thresholds ($\mathcal{O}$(1eV--10 eV)) achievable with cryogenic detectors operating at mK temperatures. ALPs~\cite{ALP1,ALP2} generalize the axion concept: they are hypothetical pseudo-scalar (spin-0) particles with properties similar to QCD axions but not tied to the strong CP problem. Unlike QCD axions, ALPs are not subject to stringent mass constraints and can span a broad range of masses and couplings.

Dark photons~\cite{DP1} are hypothetical spin-1 vector bosons predicted in extensions of the SM. They arise naturally from a hidden sector with its own $U(1)_D$ gauge symmetry, which couples to the SM photon via kinetic mixing. This mixing induces a small effective coupling between dark photons and SM charged particles, allowing dark photons to act as mediators between visible matter and the dark sector. Unlike QCD axions, dark photons do not have a fixed mass scale; depending on the model, they may decay visibly into SM particles (e.g., $e^+e^-$ pairs) or invisibly into the dark sector particles.

Both ALPs and dark photons are viable cold dark matter candidates. They can be absorbed in detector materials through the axio-electric effect and kinetic mixing (dark photons), in processes collectively known as \emph{dark absorption}~\cite{dark_absoprtion_hochberg}. This mechanism is analogous to the photoelectric effect: an incoming ALP or dark photon is absorbed by the target, ejecting an electron with excess energy. The corresponding absorption cross sections scale with the material’s photoelectric cross section and are closely tied to its complex conductivity. Assuming ALPs constitute the local dark matter halo and behave as non-relativistic particles, the expected absorption rate is~\cite{dark_absoprtion_hochberg}:
\begin{equation}\label{ALPs_rate}
R_a = \frac{\rho_{\text{DM}}}{\rho} \cdot \frac{3 m_a g_{ae}^2 \, \sigma_1(m_a)}{4 m_e^2 e^2},
\end{equation}
where $\rho$ is the detector material density, $\rho_{\text{DM}}$ is the local dark matter density, $m_a$ is the ALP mass, $m_e$ the electron mass, $e$ the elementary charge, $g_{ae}$ the axio-electric coupling constant, and $\sigma_1(m)$ the real part of the material’s conductivity.

Similarly, if dark photons constitute all of dark matter, the absorption rate is~\cite{dark_absoprtion_hochberg}:
\begin{equation} \label{DP_rate}
\begin{split}
R_V &= \frac{1}{\rho} \cdot \frac{\rho_{\text{DM}}}{m_V} \cdot \varepsilon_{\mathrm{eff}}^2 \, \sigma_1(m_V), \\
\varepsilon_{\mathrm{eff}}^2 &= \frac{\varepsilon^2 m_{V}^2}{m_{V}^2 - 2 m_{V} \sigma_2 + \sigma_2^2 + \sigma_1^2} \, ,
\end{split}
\end{equation}
where $m_V$ is the dark photon mass, $\varepsilon_{\mathrm{eff}}$ the effective kinetic mixing parameter, and $\sigma_2$ the imaginary part of the conductivity of the detector material.

\begin{figure}[h!]
    \includegraphics[width=\linewidth]{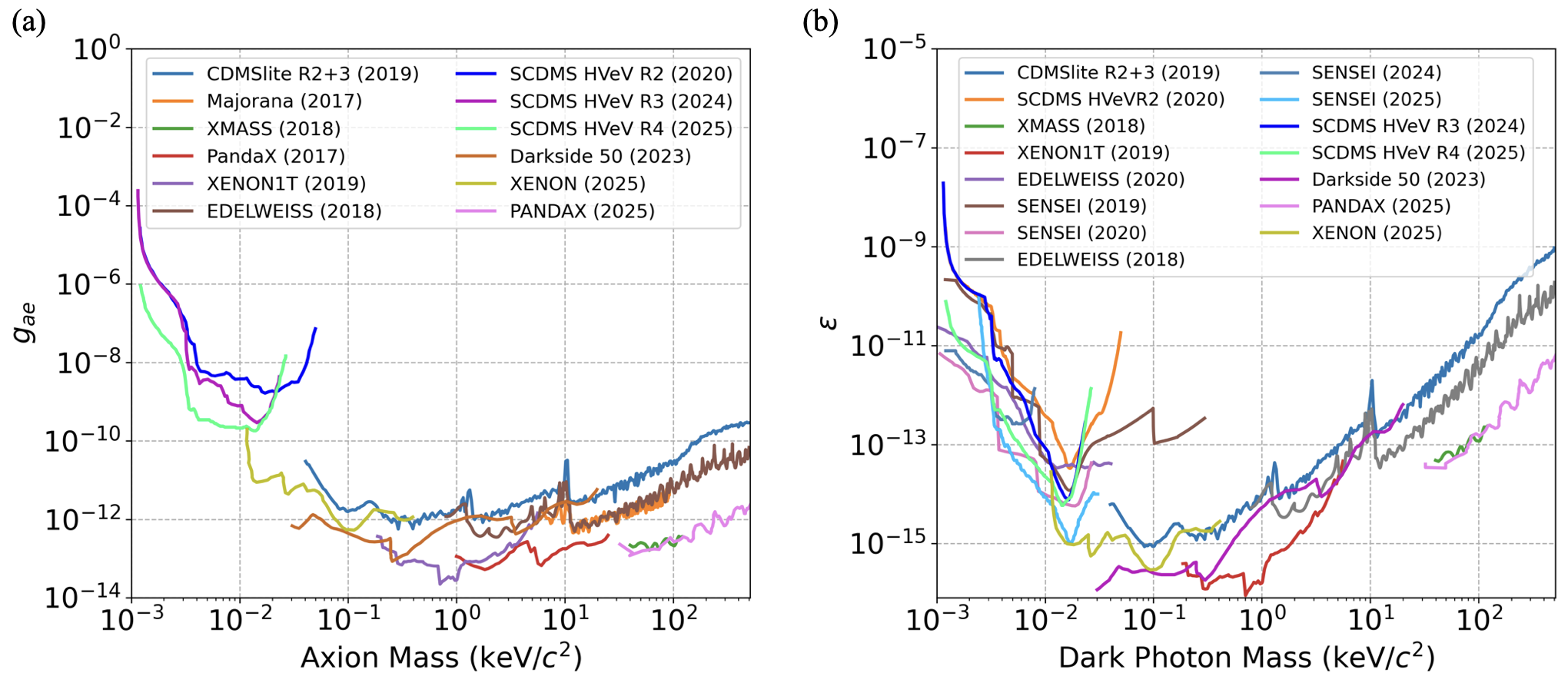}\\
    \caption{ 90\% confidence level upper limits on (a) axio-electric coupling for ALPs and  (b) the  kinetic mixing for dark photons from SuperCDMS~\cite{SCDMS_soudan_ALPs_DP,SuperCDMS_HVeV_R2_ALPs_DP,SuperCDMS_HVeV_R3_ALPs_DP,HVeVR4_2025_ALPs_DP}, EDELWEISS~\cite{EDELWEISS_III_2018_ALPs_DP}, Majorana~\cite{Majorana_2017_ALPs}, DarkSide~\cite{DarkSide_50_2023_ALPs_DP}, XENON~\cite{XENON_1T_2019_ALPs_DP,XENON_2025_ALPS_DP}, PandaX~\cite{PandaXII_2017_ALPs,PANDAX_2025_ALPS_DP}, XMASS~\cite{XMASS_2018_ALPs_DP}, and SENSEI~\cite{SENSEI_2019_DP,SENSEI_2020_DP,SENSEI:2025_PRL_DP,SENSEI:2024_1e}. Limits are obtained from publicly available data or digitized from published figures.}
    \label{axions_Dark_photons}
\end{figure}

Cryogenic detectors, such as those used in SuperCDMS Soudan and the SuperCDMS HVeV program, are suitable for ALP and dark photon searches due to their ultra-low energy thresholds ($\sim$10 eV). In analogy to photoelectric absorption, the expected signal is a monoenergetic electron recoil peak, with width determined by the detector resolution~\cite{SCDMS_soudan_ALPs_DP}. No evidence of ALPs and Dark photons has been found so far, so experiments have determined an upper limit on the strength of the respective coupling parameters. The standard procedure is to set an experimental upper bound on the event rate at a given dark-matter mass and, using the theoretical rate expressions in Eqs.~\ref{ALPs_rate} and \ref{DP_rate}, translate this bound into limits on the coupling parameters.

The leading systematic uncertainties in ALP and dark photon searches stem from detector response to low-energy electron recoils. Since the signals are expected as sharp spectral features, uncertainties in absolute energy scale, energy resolution, and low-energy calibration directly impact sensitivity. Additional uncertainties arise from photoelectric cross sections and dielectric response functions of the target material~\cite{dark_absoprtion_hochberg,SCDMS_soudan_ALPs_DP}. At the eV-scale energies, variations in the complex conductivity components ($\sigma_1$, $\sigma_2$) of materials such as Ge and Si across different measurements can introduce up to $\sim$10\% systematic uncertainty~\cite{dark_absoprtion_hochberg}. These effects are incorporated into the statistical analysis, generally leading to conservative exclusion limits. Continued improvements in calibration techniques and detector modeling are expected to significantly reduce these systematics in future searches.

A range of experiments have contributed to ALP and dark photon searches. SuperCDMS~\cite{SCDMS_soudan_ALPs_DP,SuperCDMS_HVeV_R2_ALPs_DP,SuperCDMS_HVeV_R3_ALPs_DP,HVeVR4_2025_ALPs_DP} and EDELWEISS~\cite{EDELWEISS_III_2018_ALPs_DP} use cryogenic detectors, while Majorana~\cite{Majorana_2017_ALPs} employs high-purity Ge ionization detectors. DarkSide~\cite{DarkSide_50_2023_ALPs_DP}, XENON~\cite{XENON_1T_2019_ALPs_DP}, PandaX~\cite{PandaXII_2017_ALPs}, and XMASS~\cite{XMASS_2018_ALPs_DP} utilize liquid noble detectors, and SENSEI~\cite{SENSEI_2019_DP,SENSEI_2020_DP,SENSEI:2025_PRL_DP,SENSEI:2024_1e} relies on skipper CCDs. Figure~\ref{axions_Dark_photons} compares their 90\% CL exclusion limits. Notably, the SuperCDMS HVeV program \cite{SuperCDMS_HVeV_R2_ALPs_DP,SuperCDMS_HVeV_R3_ALPs_DP,HVeVR4_2025_ALPs_DP} demonstrated that even with exposures as small as $\mathcal{O}$(1 g-day), gram-scale cryogenic detectors can produce competitive or world-leading constraints. In contrast, the SuperCDMS Soudan experiment, with exposures of $\sim$110 kg-days, achieved constraints comparable to large-scale liquid noble experiments.

Future cryogenic experiments are expected to significantly improve sensitivity. The SuperCDMS SNOLAB experiment could probe ALP-electron couplings and dark photon kinetic mixing with world-leading reach~\cite{SuperCDMS_SNOALB_SNOWMASS_2021}. The SuperCDMS HVeV effort will employ high-voltage-assisted phonon amplification to reach eV-scale thresholds, enabling absorption searches for lighter ALPs and dark photons. In parallel, CRESST-III Phase~2~\cite{CRESST_2025_projected_sensitivity} aims to extend sensitivity in the sub-keV range with improved thresholds and low-background operation. Together, these efforts will explore new parameter space and set stronger bounds on ALP and dark photon couplings.

 Limits on the axion-photon coupling, $g_{a\gamma\gamma}$, have been established by cryogenic experiments using solar- or reactor-produced ALPs~\cite{ALP3_photon_coupling}. ALPs can also be generated from photons via the Primakoff effect, Compton-like scattering, or nuclear de-excitations, then propagate to detectors where they are absorbed. While no evidence has been observed, experiments such as CDMS, MINER, EDELWEISS-II, and TEXONO have placed stringent constraints on $g_{a\gamma\gamma}$~\cite{CDMS_axion_gamma,MINER_Axion_gamma_gamma_2025,EDELWEISS-II,TEXONO_ALPs}.

Absorption-based searches for ALPs and dark photons exploit the monoenergetic nature of electron recoils, a regime where cryogenic detectors excel. Their success so far underscores the importance of precise energy calibration and knowledge of material response, as uncertainties in conductivity and photoelectric cross-sections directly limit sensitivity. Future advances will likely stem from eV-threshold phonon-amplified detectors and improved dielectric modelling, which together could tighten limits on coupling constants by orders of magnitude and extend searches into previously unexplored mass ranges.

\section{Search for Fractionally Charged Particles (FCPs)}\label{sec:FCPs}
Continuing the discussion on rare phenomena from the previous section, we now explore another rare interaction search in direct detection experiments: the search for unconfined FCPs, which have gained significant attention in recent years. All charged particles discovered so far carry charges that are integral multiples of the electron charge. Quarks, though confined, exhibit fractional charges of $\pm \tfrac{1}{3}e$ or $\pm \tfrac{2}{3}e$, where $e$ is the charge of an electron, but in unconfined states all observed particles carry integer multiples of $e$. The underlying reason for charge quantization remains one of the open questions of modern physics. While theoretical arguments exist---notably those of Dirac and others~\cite{proc:dirac,pdg}---experimental confirmation is lacking. At the same time, many extensions of the Standard Model predict the existence of unconfined FCPs, with charges $q=\pm f e$ where $0<f<1$~\cite{paper:fcp-review}. In addition, certain dark matter models postulate milli-charged dark matter particles, whose small fractional charges would allow them to evade existing constraints~\cite{paper:mqDM}. Non-relativistic FCPs have even been suggested as explanations for the annual modulation signals reported by DAMA/LIBRA~\cite{paper:DAMA} and CoGeNT~\cite{paper:CoGeNT}, as well as in related theoretical interpretations~\cite{PhysRevD.90.12130,FOOT201561}. Since the mean energy loss per unit length, or the stopping power, is proportional to $q^2$, free FCPs with small electric charges are also referred to as Lightly Ionizing Particles (LIPs) in the literature. For the rest of this section, we will use these terms interchangeably.

A wide variety of experimental strategies have been pursued to search for FCPs, including fixed-target accelerators~\cite{paper:Aubert,paper:Bergsma,paper:Golowich,paper:Huntrap,paper:Ghosh,paper:Prinz,paper:Soper}, collider experiments~\cite{paper:Abe,paper:Buskulic,paper:Akers,paper:Abreu,paper:Acosta,paper:Abbiendi,paper:Jaeckel,paper:Chatrchyan}, reactor-based searches~\cite{paper:Gninenko,paper:Chen,paper:Singh}, bulk matter studies~\cite{paper:Joyce,paper:Halyo,paper:Lee,paper:Kim,paper:Larue,paper:Marinelli,paper:Smith,paper:Smith2,paper:Jones,paper:Homer}, and direct-detection experiments~\cite{limit:MACRO-LIPs,limit:Kamiokande-LIPs,limit:LSD-LIPs,limit:CDMSII-LIPs,limit:Majorana-LIPs,limit:TEXONO-LIPs,limit:CDMSlite-LIPs,limit:CUORE}. Direct detection is particularly advantageous for two reasons: (i) sensitivity to FCPs with masses kinematically inaccessible to collider or reactor experiments, and (ii) the ability to detect cosmogenic FCPs incident on Earth. However, the reach of such searches is limited by detector exposure and the flux of FCPs crossing the instrument. 

The direct detection of FCPs involves searching for events consistent with FCP signatures but not with backgrounds. Since the expected number of signal events is proportional to the intensity, the upper limit on the intensity at a 90\,\% confidence level ($I_v^{90}$) is computed in the absence of a significant excess. The general formula for $I_v^{90}$ is given as:
\begin{equation}
I_v^{90} = \frac{N^{90}}{A\times\tau\times\epsilon\times\Omega},
\end{equation}
where A is the cross-sectional area of the detector, $\tau$ is the time over which the search is carried out, $\epsilon$ is the detection efficiency, and $\Omega$ is the solid angle, or geometrical acceptance of the detector. $N_{90}$ is computed via various statistical techniques, including classical Poisson method~\cite{ParticleDataGroup:2024cfk} (e.g., Ref.~\cite{limit:CDMSII-LIPs}),  Feldman–Cousins method~\cite{FC-PhysRevD.57.387} (e.g., Ref.~\cite{limit:Majorana-LIPs}), and the Optimum Interval ~\cite{yelllin-PhysRevD.66.032005} approach (e.g., Ref.~\cite{limit:CDMSlite-LIPs}). $\epsilon$ is determined using Monte Carlo simulations, accounting for FCP interaction probability in the detector, and trigger and data selection efficiencies.

\begin{figure}[h]
    \includegraphics[width=0.49\linewidth]{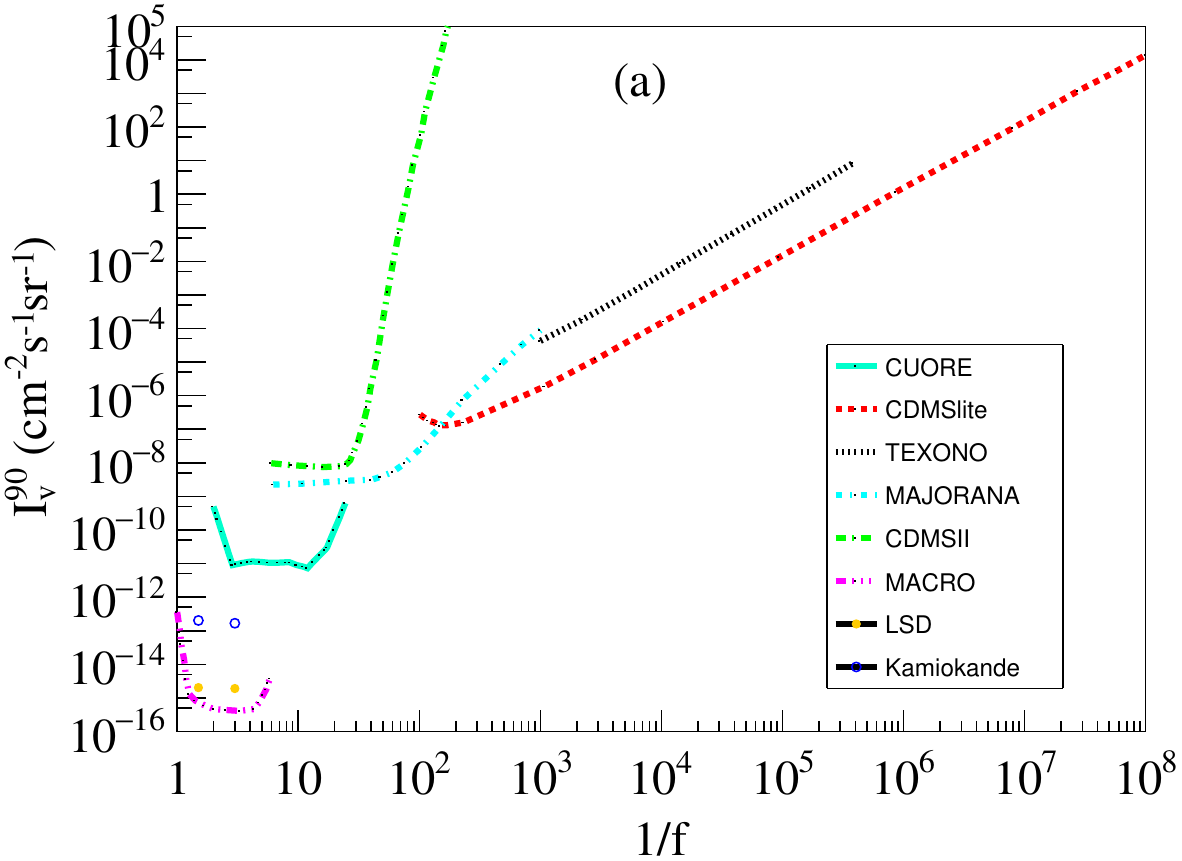}
    \includegraphics[width=0.49\linewidth]{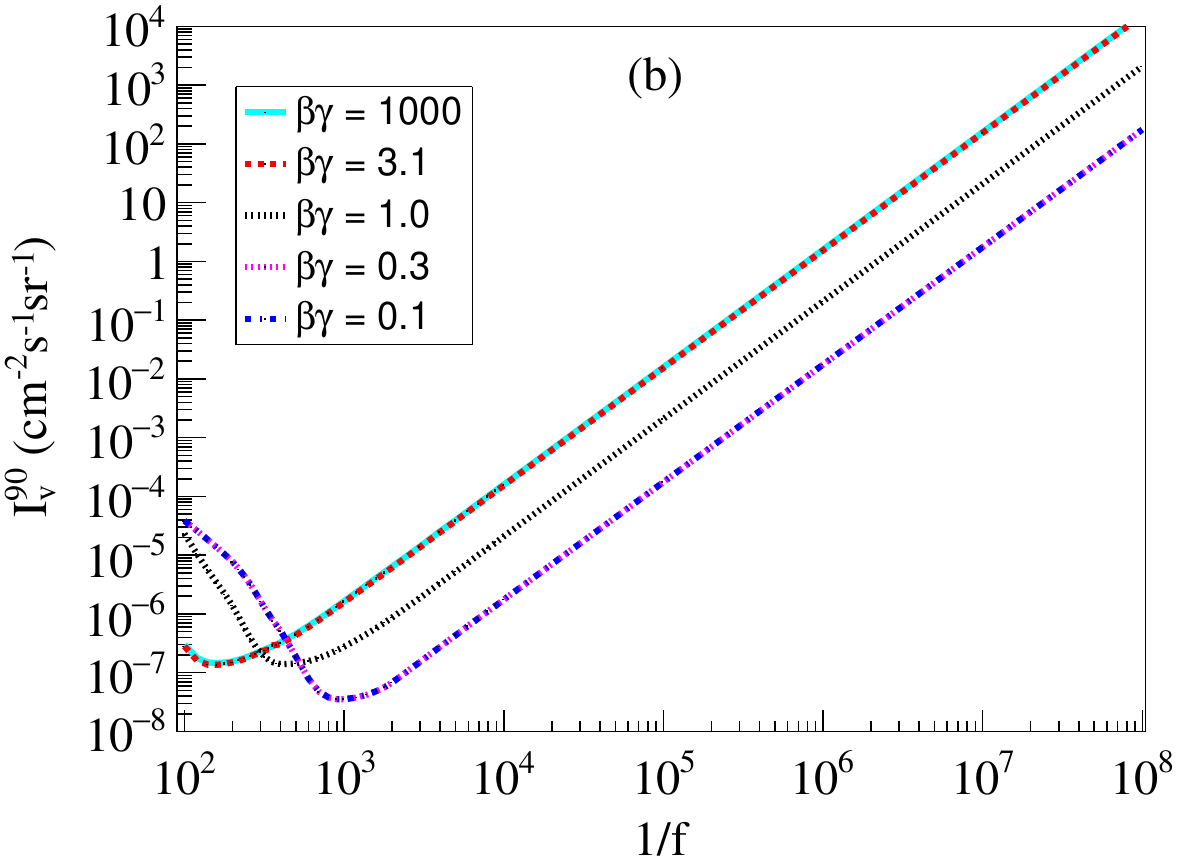}   
    \caption{(a) 90\% confidence level upper limits ($I_v^{90}$) on the intensity of lightly ionizing particles (LIPs) from CUORE~\cite{limit:CUORE}, CDMSlite~\cite{limit:CDMSlite-LIPs}, TEXONO~\cite{limit:TEXONO-LIPs}, Majorana~\cite{limit:Majorana-LIPs}, CDMS-II~\cite{limit:CDMSII-LIPs}, MACRO~\cite{limit:MACRO-LIPs}, LSD~\cite{limit:LSD-LIPs}, and Kamiokande~\cite{limit:Kamiokande-LIPs}, assuming an isotropic distribution of minimum-ionizing LIPs. (b) 90\% confidence upper limits from CDMSlite~\cite{limit:CDMSlite-LIPs} on LIP intensity as a function of charge fraction $f$ for different values of $\beta\gamma$. Curves for $\beta\gamma \gtrsim 10^3$ coincide and are represented by a single line.}
    \label{fig:fcp_exclusion}
\end{figure}

Direct searches for FCPs include MACRO~\cite{limit:MACRO-LIPs}, Kamiokande~\cite{limit:Kamiokande-LIPs}, LSD~\cite{limit:LSD-LIPs}, CDMS-II~\cite{limit:CDMSII-LIPs}, Majorana~\cite{limit:Majorana-LIPs}, TEXONO~\cite{limit:TEXONO-LIPs}, CDMSlite~\cite{limit:CDMSlite-LIPs}, and CUORE~\cite{limit:CUORE}. Collectively, these experiments exclude large regions of the FCP intensity--charge parameter space. Figure~\ref{fig:fcp_exclusion} summarizes the limits: most experiments probe minimum-ionizing LIPs [Fig.~\ref{fig:fcp_exclusion}(a)], while CDMSlite has uniquely excluded broad regions of velocity parameter space, spanning $\beta\gamma$\footnote{Here $\beta=v/c$, $\gamma=1/\sqrt{1-\beta^2}$ with $v$ being the velocity of LIPs.} from 0.1 to $10^{6}$ [Fig.~\ref{fig:fcp_exclusion}(b)].

The treatment of systematic uncertainties differs significantly among experiments. For Kamiokande~\cite{limit:Kamiokande-LIPs}, LSD~\cite{limit:LSD-LIPs}, and MACRO~\cite{limit:MACRO-LIPs}, no explicit uncertainty estimates are provided. For CDMS-II~\cite{limit:CDMSII-LIPs}, the dominant contributions come from detector thickness, analysis efficiency, and calibration, amounting to systematic uncertainties below 25\%. In Majorana~\cite{limit:Majorana-LIPs}, uncertainties arise from fiducial thickness and energy resolution, while for CDMSlite~\cite{limit:CDMSlite-LIPs} they include data-selection efficiency, threshold uncertainties, and modeling of the signal response, leading to combined uncertainties of $\sim$37\%. CUORE~\cite{limit:CUORE} quantifies its uncertainty bands through $\pm 1\sigma$ and $\pm 2\sigma$ expected limits under the background-only hypothesis. In general, the uncertainties shown in Fig.~\ref{fig:fcp_exclusion} are smaller than the width of the exclusion curves themselves.

The ongoing development of more sensitive detectors and refined experimental techniques will continue to push the boundaries of LIP searches, providing new constraints and insights into their existence. The SuperCDMS experiment is preparing to enhance its sensitivity to LIPs with an upcoming upgrade at SNOLAB~\cite{Zatschler:2024VW}. Meanwhile, the CUORE experiment is refining its analysis and processing methods to lower energy thresholds and improve efficiency, allowing it to detect smaller fractional charges than those studied in previous research~\cite{limit:CUORE}. By employing similar search strategies, enhancing detector segmentation, and incorporating additional information from the dual readout of heat and light signatures in the CUORE Upgrade with Particle Identification (CUPID), the experiment aims to extend its sensitivity into previously unexplored regions of the LIPs parameter space~\cite{limit:CUORE,cupid-cdr}.

\section{Search for Coherent Elastic Neutrino-Nucleus Scattering (CE$\nu$NS) \label{sec:CEvNS}}
Similar to DM and LIPs, CE$\nu$NS is a rare low-energy process. Unlike DM and LIPs, which involve new physics beyond the SM, CE$\nu$NS is a theoretically well-understood SM prediction. The process was first proposed by Daniel Z. Freedman in 1974~\cite{freedman}. The defining features of this interaction are embedded in the name: the interaction is \emph{coherent} because, at low momentum $\vec{q}$ transfer ($|\vec{q}| \lesssim 1/R_N$), the neutrino’s de Broglie wavelength is comparable to or larger than the nuclear radius $R_N$, leading all nucleons to recoil in phase with the transferred energy. The scattering is \emph{elastic}, meaning the nucleus neither breaks up nor becomes excited, so kinetic energy is conserved. Under these conditions, the neutrino interacts with the entire nucleus, resulting in a cross section that scales approximately with the square of the neutron number. CE$\nu$NS is a flavour-independent neutral-current process mediated by the exchange of a $Z$ boson, with its SM differential cross section given by~\cite{CEvNS_crossSection}:

\begin{equation}
    \frac{d\sigma_{\nu N}}{dT}(E_{\nu}, T)=\frac{G^2_FM}{4\pi} \Big( 1-\frac{MT}{2E^2_{\nu}}\Big) Q^2_WF^2(q^2) 
    \label{eq:CEvNS_crossSection}
\end{equation}
where $E_{\nu}$ is the neutrino energy, $T$ the nuclear recoil energy, $M$ the nuclear mass, $G_F$ the Fermi constant, $Q_W$ the weak nuclear charge, and $F(q^2)$ the nuclear form factor. Experimentally, the measured recoil spectrum is compared with predictions from Eq.~\ref{eq:CEvNS_crossSection}. Observing CE$\nu$NS provides both a test of SM predictions and a sensitive probe of new physics. Within the SM, CE$\nu$NS allows precise measurements of the weak mixing angle~\cite{weak_mixing_angle} and neutron distributions in nuclei~\cite{neutron_form_factor}. Beyond the SM, it constrains neutrino electromagnetic properties~\cite{neutrino_electromagnetic_properties} and probes the existence of new mediators~\cite{light_mediator}, among other effects. As discussed in Section~\ref{sec:WIMP}, CE$\nu$NS is also an irreducible background for DM searches, underscoring the importance of accurate modeling.

The first observation of CE$\nu$NS came in 2017 by the COHERENT collaboration using a 14.6 kg CsI[Na] scintillator at the Spallation Neutron Source (SNS)~\cite{COHERENT}. These neutrinos, produced via $\pi^+$ decay-at-rest, have mean energies of about 30 MeV and generate nuclear recoils of a few to tens of keV, which are detectable with standard scintillators. However, at this energy, partial loss of coherence could occur.  Reactor-based CE$\nu$NS experiments, in contrast, use $\bar{\nu}_e$ with energies up to $\sim$10 MeV~\cite{coherency}. These lower energies preserve full coherence~\cite{coherency}, but yield recoil energies of the order of a few keV or less, requiring detectors with ultra-low thresholds and excellent noise suppression. Cryogenic detectors, with their high resolution and sensitivity to sub-keV signals, are ideally suited to this regime. TEXONO and CONUS+ employ ionization-based HPGe detectors (see Section~\ref{sec:ionization_based}), while NUCLEUS and Ricochet deploy cryogenic calorimeters with transition-edge sensors (TES) (see Section~\ref{phonon_detector}) to detect CE$\nu$NS.

As with DM searches, CE$\nu$NS sensitivities are limited by systematic uncertainties. The dominant source of uncertainty is the nuclear recoil quenching factor, which governs the signal conversion in both HPGe (see Section~\ref{sec:WIMP}) and scintillating calorimeters.

Additional uncertainties stem from reactor antineutrino flux estimates~\cite{CEvNS_flux_systematic}, and from backgrounds due to reactor-produced gammas and neutrons. Further contributions include calibration, data processing, and event selection. Together, these systematics must be tightly controlled to isolate the CE$\nu$NS signal.

\begin{figure}[h]
    \centering
    \includegraphics[width=1\linewidth]{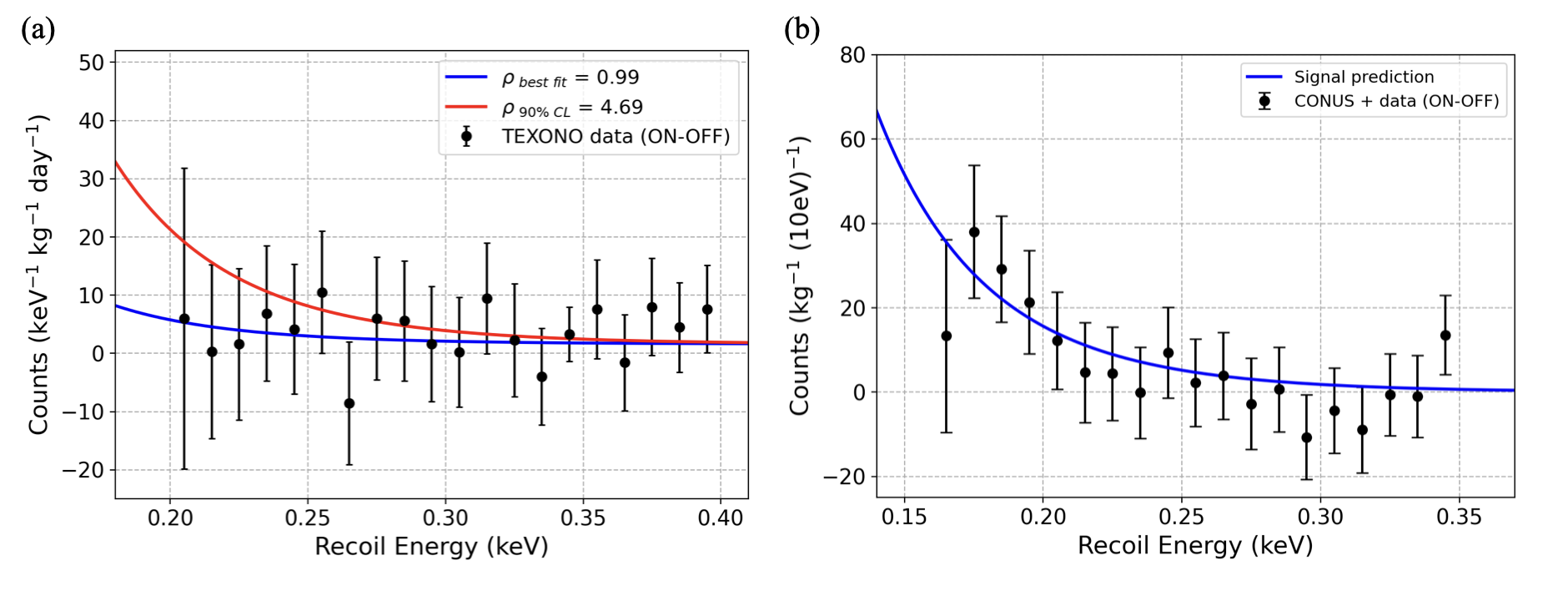}
    \caption{\label{fig:texono_conus} (a) TEXONO results~\cite{TEXONOresult}, showing digitized data points compared with model predictions. (b) CONUS+ results~\cite{COUNSplus} and comparison with model prediction, reporting the first evidence of CE$\nu$NS with reactor antineutrinos.}
\end{figure}

The TEXONO program at the Kuo-Sheng Reactor Neutrino Laboratory (KSNL) searches for CE$\nu$NS using PPC-based HPGe detectors (see Section~\ref{sec:ionization_based}). Its latest analysis used 242 (357) kg·days of reactor ON (OFF) data collected between 2020 and 2023 with a 200 eV$_\mathrm{ee}$ threshold. TEXONO set a 90\% C.L. upper limit of 4.69 on the ratio $\rho$ of the measured to the Standard Model predicted cross section [see Fig.~\ref{fig:texono_conus}(a)], assuming an ionization quenching factor of $k=0.162$~\cite{TEXONOresult}.

CONUS+ operates HPGe detectors with thresholds as low as 160--180 eV$_\mathrm{ee}$. The most recent results  [Fig.~\ref{fig:texono_conus}(b)] report the first evidence for reactor-based CE$\nu$NS at a significance of 3.7$\sigma$~\cite{COUNSplus}. The observed Ge quenching factor is consistent with Lindhard theory ($k=0.162$~\cite{lindhard_factor}), ruling out deviations reported in some earlier studies~\cite{lindhard_factor_deviation}.

The MINER experiment carried out its first search for CE$\nu$NS at a baseline distance of approximately 4~m from a 1~MW$_{\text{th}}$ TRIGA reactor~\cite{MINER_CEvNS}. A 72~g sapphire detector with phonon sensors (\(\mathrm{Al_{2}O_{3}}\)) and an energy threshold of 250~eV$_{\text{nr}}$ was used. In this measurement, the experiment set an upper limit on the parameter $\rho$ of 2524, with the data indicating background dominance over the expected signal. The next phase of the experiment is planned to be conducted at the High Flux Isotope Reactor (HFIR) at Oak Ridge National Laboratory~\cite{reactor_HFIR}, which operates at a higher power of 85~MW$_{\text{th}}$ with a baseline distance of approximately 5~m. With this upgrade, along with an improved shielding configuration, MINER is expected to potentially detect CE$\nu$NS or set more stringent limits on $\rho$.


Ricochet is installing a kilogram-scale cryogenic payload at the Institut Laue-Langevin (ILL), comprising CryoCube and Q-Array technologies~\cite{Ricochet_cryocube}: with 18--27 Ge and 9 Zn crystals, each $\approx$~38 g. CryoCubes measure phonons with NTD sensors and ionization with HEMT amplifiers. A Ge prototype demonstrated 30 eV$_\mathrm{ee}$ resolution, with a target energy resolution of 20 eV$_\mathrm{ee}$~\cite{Ricochet}. Following installation in 2024, initial tests achieved 40 eV$_\mathrm{ee}$ ionization resolution and 50--80 eV$_\mathrm{nr}$ phonon resolution, marking readiness for the upcoming science run~\cite{Ricochet_comissioning}.


NUCLEUS employs gram-scale cryogenic calorimeters of Al$_2$O$_3$ and CaWO$_4$, each equipped with TES sensors. A 0.5 g Al$_2$O$_3$ prototype reached a 19.7 $\pm$ 0.9 eV$_\mathrm{nr}$ threshold~\cite{NUCLEUS_prototype}. Its multi-target design uses nine CaWO$_4$ (total $\approx$ 6 g) and nine Al$_2$O$_3$ (total $\approx$ 4 g) cubes to measure CE$\nu$NS and backgrounds simultaneously~\cite{NUCLEUS_bkg}. In its 2024 commissioning phase, two detectors achieved resolutions of 5.5 eV$_\mathrm{nr}$ (Al$_2$O$_3$) and 6.5 eV$_\mathrm{nr}$ approximately (CaWO$_4$), validating background characterization and detector stability. A technical run is planned for 2026~\cite{NUCLEUS_comissioning}.


\begin{table}[h]
\centering
\caption{Reactor-based cryogenic CE$\nu$NS experiments.}
\label{table:cenns_experiments}
\resizebox{\linewidth}{!}{
\begin{tabular}{llcccc}
\hline
\textbf{Experiment} & \textbf{Detector Material}& \textbf{Mass}  & \textbf{Threshold }   & \textbf{Reactor power}  &   \textbf{Baseline}\\
\hline

CONUS$+$ \cite{COUNSplus}    & HPGe   &  3.74 kg  & (160 - 180) eV$_\mathrm{ee}$     &  3.6 GW$_\mathrm{th}$ & 20.7 m\\

TEXONO \cite{TEXONOresult}     & HPGe   & 1.4 kg   &200 eV$_\mathrm{ee}$  & 2.9 GW$_\mathrm{th}$ & 28 m\\
nuGEN \cite{nuGEN}     & HPGe   & 1.4 kg   &290 eV$_\mathrm{ee}$  &  3.1 GW$_\mathrm{th}$   & 11.8 m\\

MINER \cite{MINER-sapphire, miner, MINER_CEvNS} & Al$_2$O$_3$     & 72 g
& 250  eV$_\mathrm{nr}$     &  1 MW$_\mathrm{th}$ & 4 m\\
Ricochet \cite{Ricochet}   &Ge, Zn, Al  &680 g
& 150 eV$_\mathrm{ee}$  & 58 MW$_\mathrm{th}$ & 8.8 m\\
NUCLEUS \cite{NUCLEUS_bkg,NUCLEUS_prototype}   &CaWO$_4$, Al$_2$O$_3$     &10 g
& 20 eV$_\mathrm{nr}$  & $2\times4.25$ GW$_\mathrm{th}$ & 72 m \& 102 m\\
\hline
\end{tabular}
}
\end{table}

Several cryogenic experiments have now reached the sensitivity required to explore CE$\nu$NS at reactors, making steady progress by lowering thresholds, suppressing backgrounds, and refining calibration. As summarized in Table~\ref{table:cenns_experiments}, these efforts highlight the growing potential of cryogenic detectors to enable precision CE$\nu$NS measurements, with wide-ranging implications for both SM tests and searches for new physics.

The first evidence of reactor CE$\nu$NS marks a milestone for cryogenic detectors, yet it also exposes key challenges. Quenching-factor uncertainties, reactor flux modelling, and background subtraction now dominate experimental systematics. Phonon-based calorimeters have demonstrated the resolution needed to probe differential recoil spectra and electroweak parameters, but scaling these systems while maintaining noise suppression remains non-trivial.

Future efforts will likely focus on multi-target deployments, improved flux characterisation, and combined reactor- and accelerator-based measurements, enabling precision tests of the Standard Model and competitive sensitivity to new physics.

\section{Search for Neutrinoless Double Beta Decay ($0\nu\beta\beta$)\label{sec:NDBD}}
Continuing the discussion of rare-event phenomena, $0\nu\beta\beta$ represents a particularly significant case where double beta decay itself is a rare second-order weak nuclear process in which two neutrons in a nucleus simultaneously convert into two protons. The Standard Model allowed mode, known as two-neutrino double beta decay ($2\nu\beta\beta$), has been observed in several isotopes~\cite{doublebetadecay}:
\begin{equation}
\mathrm{(A, Z) \rightarrow (A, Z+2) + 2 e^{-}} + 2\bar{\nu}_{e}
\end{equation}
This process conserves lepton number, as two antineutrinos are emitted along with the electrons. In 1939, W. H. Furry proposed a neutrinoless mode of double beta decay ($0\nu\beta\beta$), in which no neutrinos are emitted~\cite{Furry1939}:
\begin{equation}
\mathrm{(A, Z) \rightarrow (A, Z+2) + 2 e^{-}}
\end{equation}

Observation of $0\nu\beta\beta$ would establish lepton number violation and confirm that neutrinos are Majorana particles—fermions that are their own antiparticles~\cite{Schechter1981}. Such a discovery would constitute direct evidence of physics beyond the SM, while also providing insights into the absolute neutrino mass scale, the mechanism of neutrino mass generation, and the origin of the matter–antimatter asymmetry of the Universe~\cite{FUKUGITA}.

Although $\sim$35 naturally occurring even-even isotopes are energetically allowed to undergo double beta decay~\cite{Barabash2015}, only a subset, including $^{76}$Ge, $^{130}$Te, $^{100}$Mo, and $^{136}$Xe, are practical candidates for experimental searches owing to their high Q-values (above most natural backgrounds), chemical compatibility with detector technologies, and isotopic availability. The primary observable is the half-life $T_{1/2}^{0\nu}$, which is related to the effective Majorana neutrino mass $\langle m_{\beta\beta} \rangle$ via
\begin{equation}\label{eq:NDBD1}
\left(T_{1/2}^{0\nu}\right)^{-1} = G^{0\nu} g_A^{4} |M^{0\nu}|^{2} \langle m_{\beta\beta} \rangle^{2},
\end{equation}
where $G^{0\nu}$ is the phase-space factor, $g_A$ the axial-vector coupling constant, and $M^{0\nu}$ the nuclear matrix element reflecting the underlying nuclear structure~\cite{KlapdorKleingrothaus2000}. Measuring or constraining $T_{1/2}^{0\nu}$ therefore directly probes $\langle m_{\beta\beta} \rangle$.

\begin{figure}[tbh]
\centering
\includegraphics[width=0.49\linewidth]{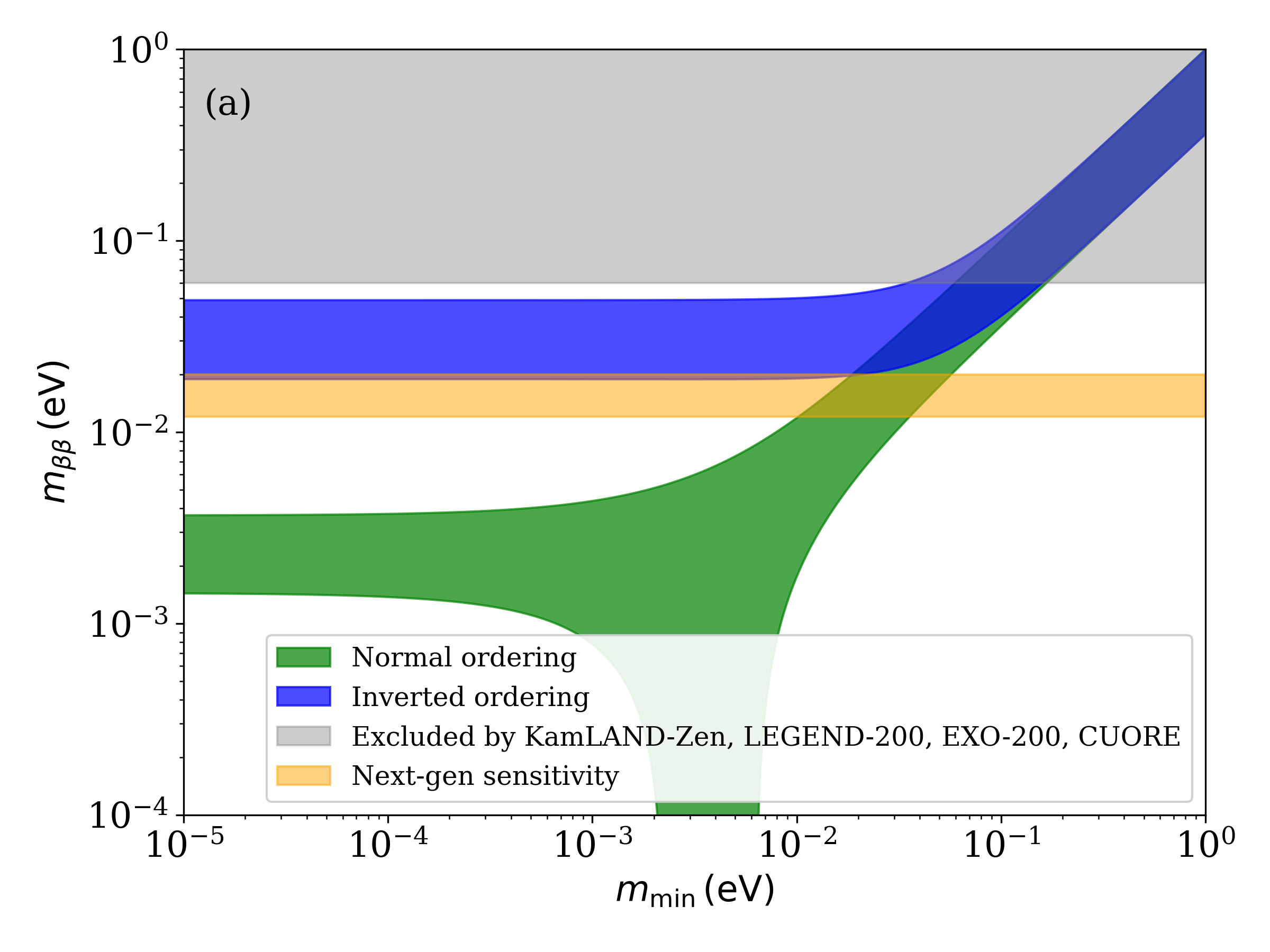}
\includegraphics[width=0.49\linewidth]{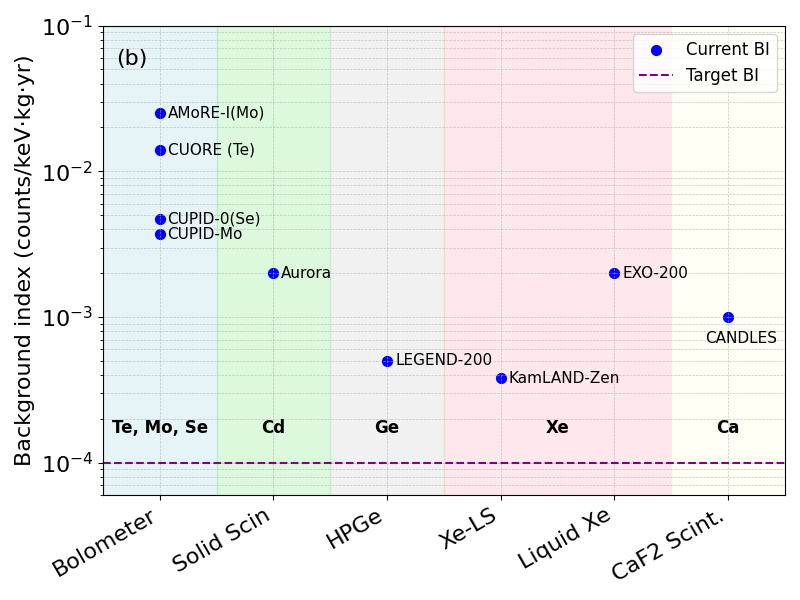}
\caption{(a) Regions of the effective Majorana mass, $m_{\beta\beta}$ that are allowed by neutrino oscillation data, shown as a function of the lightest neutrino mass, $m_\mathrm{min}$. The green band corresponds to normal ordering ($m_1<m_2\ll m_3$) while the blue band represents inverted ordering ($m_3\ll m_1\lesssim m_2$) \cite{DellOro2016}. The gray shaded region represents the sensitivity of current experiments while the yellow band shows the projected reach of next-generation searches. Next-generation experiments will be able to fully test the inverted ordering scenario. (b) Current background indices (BI) of leading $0\nu\beta\beta$ experiments using different detector technologies. The horizontal dashed line marks the target BI needed to explore the inverted ordering regime. Although BI values appear comparable across technologies, the effective background depends strongly on the detector mass scale.}
\label{fig:NDBD_sensi}
\end{figure}

The hallmark signature of $0\nu\beta\beta$ is a monoenergetic peak at the decay Q-value, equal to the summed energy of the two emitted electrons. By contrast, the $2\nu\beta\beta$ spectrum is continuous up to the same endpoint. Distinguishing $0\nu\beta\beta$ from backgrounds requires: (i) excellent energy resolution to suppress the $2\nu\beta\beta$ tail, (ii) ultra-low backgrounds ($<10^{-4}$ counts/(keV·kg·yr)), (iii) large active mass for sufficient statistics, and (iv) high detection efficiency. Current experiments like KamLAND-Zen \cite{abe2024}, LEGEND-200 \cite{LEGEND2025}, EXO-200 \cite{EXO2002019}, CUORE \cite{Campani2024} have achieved half-life sensitivities in the range of $10^{25}$–$10^{26}$ yr, which correspond to limits on the $\langle m_{\beta\beta} \rangle$ at the 50–200 meV scale, as illustrated by the gray shaded region in Fig.~\ref{fig:NDBD_sensi}(a). However, to fully explore the inverted neutrino mass hierarchy and approach the normal hierarchy regime (as indicated by the yellow band in Fig.~\ref{fig:NDBD_sensi}(a) respectively), next-generation detectors must reach sensitivities of $10^{27}$–$10^{28}$ yr.

Bolometric detectors satisfy many of these requirements by measuring temperature rises in dielectric crystals operated at mK temperatures. They routinely achieve resolutions better than 5 keV at 2–3 MeV, and, when combined with scintillation or ionization channels, can discriminate between $\alpha$, $\beta/\gamma$, and neutron interactions~\cite{Poda2017}. A detailed discussion of their detection principle can be found in Section~\ref{sec:cryogenic_detector_technology}. Despite these advantages, systematic uncertainties remain. Calibration and resolution near the Q-value contribute 1–3\% uncertainty \cite{CUORE:2014xzu}, background modeling (including surface $\alpha$ events and cosmogenic activation) can shift inferred limits by 5–10\% \cite{CUORE:2024fak}, while detection efficiency contributes 2–4\% \cite{GERDA2020}. Isotopic enrichment fraction and fiducial mass determination add uncertainties at the $\sim$1\% level. Current and next-generation experiments mitigate these with frequent calibrations, multi-channel readout for particle ID, precision isotopic assays, and underground storage of materials to minimize cosmogenics. With such strategies, tonne-scale detectors aim to reduce total systematics below $\sim$3–5\%, keeping statistical sensitivity as the dominant factor.

\textbf{CUORE:} The Cryogenic Underground Observatory for Rare Events (CUORE) at Gran Sasso is the first ton-scale bolometric experiment dedicated to $0\nu\beta\beta$. It employs 988 TeO$_2$ crystals (742 kg total, including 206 kg $^{130}$Te), achieving 7.7 keV (FWHM) resolution at the $^{130}$Te Q-value of 2527 keV, with a background index of $\sim 1.4\times 10^{-2}$ counts/(keV·kg·yr). Based on $\sim$1 ton·yr exposure, CUORE reported $T_{1/2}^{0\nu} > 2.2 \times 10^{25}$ yr (90\% C.L.)~\cite{CUORE2021}, recently updated to $3.3 \times 10^{25}$ yr with 2023 kg·yr exposure~\cite{Campani2024}, corresponding to $\langle m_{\beta\beta} \rangle < 75$–255 meV.

\textbf{CUPID:} Building on CUORE’s infrastructure, CUPID (CUORE Upgrade with Particle ID) employs scintillating bolometers of enriched $^{100}$Mo in Li$_2$MoO$_4$, enabling $\alpha/\beta$ discrimination. Prototypes CUPID-0 ($^{82}$Se)~\cite{CUPID2022} and CUPID-Mo ($^{100}$Mo)~\cite{Augier2022} demonstrated background indices below $10^{-3}$ counts/(keV·kg·yr) and resolutions $\lesssim$6 keV. CUPID aims for half-life sensitivity beyond $10^{27}$ yr, probing $\langle m_{\beta\beta} \rangle$ down to $\sim$10 meV~\cite{CUPID2025,CUPID2025_2}.
\begin{table}[h!]
\centering
\caption{Selected $0\nu\beta\beta$ decay experiments, their isotopes, techniques, and current best half-life limits.}
\label{table:ndbd}
\resizebox{\linewidth}{!}{%
\begin{tabular}{llclc}
\hline
\textbf{Experiment} & \textbf{Isotope} & \textbf{Q-value} & \textbf{Technology} & \textbf{T$_{1/2}^{0\nu}$ Limit} \\
 &  & \textbf{(keV)} &  & \textbf{(yr)} \\
\hline
CUORE \cite{Campani2024}           & $^{130}$Te & 2527 & Bolometer (TeO$_2$) & $3.3 \times 10^{25}$ \\
CUPID-0 \cite{CUPID2022}        & $^{82}$Se  & 2998 & Scint. Bolometer    & $4.7 \times 10^{24}$ \\
CUPID-Mo \cite{Augier2022}       & $^{100}$Mo & 3035 & Scint. Bolometer    & $1.8 \times 10^{24}$ \\
AMoRE-I \cite{AMoRE2024PRL}          & $^{100}$Mo & 3035 & Bolometer + MMC     & $> 2.9 \times 10^{24}$ \\
Aurora \cite{Barabash2018}         & $^{116}$Cd & 2813 & Solid Scint.        & $>2.2 \times 10^{23}$ \\
TIN.TIN (R \& D) \cite{TinTin2020}  & $^{124}$Sn & 2292 & Bolometer (Sn)      & ---                  \\
LEGEND-200 \cite{LEGEND2025}      & $^{76}$Ge  & 2039 & HPGe (Low Bkg)      & $2.8 \times 10^{26}$           \\
KamLAND-Zen \cite{abe2024}    & $^{136}$Xe & 2459 & Liquid Scintillator & $> 3.8 \times 10^{26}$ \\
EXO-200 \cite{EXO2002019}        & $^{136}$Xe & 2459 & LXe TPC             & $3.5 \times 10^{25}$ \\
PandaX-4T \cite{PandaX2024}        & $^{136}$Xe & 2459 & LXe TPC             & $2.1 \times 10^{24}$ \\
nEXO (proj.) \cite{nEXO2017}   & $^{136}$Xe & 2459 & LXe TPC             & $>10^{28}$ (goal)    \\
SNO+ \cite{SNO2021}           & $^{130}$Te & 2527 & Liquid Scintillator & $>10^{26}$ (goal)                  \\
CANDLES \cite{CANDLES2020}        & $^{48}$Ca  & 4263 & CaF$_2$ Scint.      & $6.2 \times 10^{22}$ \\
SuperNEMO \cite{Povinec2017}      & $^{82}$Se  & 2998 & Tracker-Calorimeter & $>10^{26}$ (goal)                \\
\hline
\end{tabular}}
\end{table}

\textbf{AMoRE:} The Advanced Mo-based Rare Process Experiment (AMoRE) searches for $0\nu\beta\beta$ with $^{100}$Mo in CaMoO$_4$ and Li$_2$MoO$_4$ crystals, read out with metallic magnetic calorimeters (MMCs)~\cite{AMoRE2015}. AMoRE-I, using $\sim$100 kg of enriched crystals, achieved $T_{1/2}^{0\nu} > 2.9 \times 10^{24}$ yr~\cite{AMoRE2024PRL}. AMoRE-II will deploy $\sim$180 kg with improved shielding, targeting $T_{1/2}^{0\nu} > 6 \times 10^{26}$ yr ($\langle m_{\beta\beta} \rangle \sim$10–25 meV)~\cite{AMoRE2024_2}.

\textbf{TIN.TIN:} The India-based experiment Tin.Tin proposes Tin ($^{124}$Sn) bolometers to search for $0\nu\beta\beta$ ~\cite{TinTin2020}. A 27-element array of cryogenic Sn crystals ($3\times 3 \times 3$ cm$^3$ each) is planned, aiming for $\sim$10 keV resolution and $\alpha/\beta$ discrimination via phonon-light readout. Currently in R \& D phase, TIN.TIN represents India’s major effort in $0\nu\beta\beta$ searches.
 
\textbf{Complementary Experiments:} Non-bolometric searches provide additional coverage. KamLAND-Zen, dissolving $^{136}$Xe in a kiloton-scale scintillator, sets the strongest current limit: $T_{1/2}^{0\nu} > 3.8 \times 10^{26}$ yr~\cite{abe2024}. nEXO will deploy 5 tons of LXe in a TPC, aiming for $>10^{28}$ yr sensitivity~\cite{nEXO2017}. LEGEND, using enriched $^{76}$Ge in HPGe detectors, has already surpassed $10^{26}$ yr with LEGEND-200~\cite{LEGEND2025}, with LEGEND-1000 projected to approach $10^{28}$ yr~\cite{Calgaro2024}. Other efforts include PandaX-4T (3.7 tonne of LXe in a TPC), SNO+ (3.9 tonne of $^{130}$Te in liquid scintillator with total 780 tonne mass), CANDLES ($^{48}$Ca in 96 CaF$_2$ crystals with total mass of 305 kg), and SuperNEMO (100 kg of $^{82}$Se with tracker-calorimeter)~\cite{PandaX2024, CANDLES2020,Povinec2017}. 

A comparative overview of leading and next-generation $0\nu\beta\beta$ decay searches, including their isotopes, detector technologies, and current half-life limits, is presented in Table~\ref{table:ndbd}. While Figure~\ref{fig:NDBD_sensi}(b) illustrates the BI achieved across these experiments, scintillating bolometers with simultaneous heat–light readout stand out by providing powerful fiducialization and $\alpha/\beta$ particle-identification. This combination breaks background degeneracies that limit bare TeO$_2$ crystals, enabling BI below $\sim 10^{-3}$ counts/(keV·kg·yr)—the threshold necessary to fully probe the inverted-hierarchy region envisioned for CUPID-like next-generation detectors. Such advances highlight how detector innovation directly translates into the sensitivity required to explore the most compelling regions of the neutrino mass parameter space.

Cross-validation across isotopes and techniques will be essential to confirm any potential signal and establish the Majorana nature of neutrinos. Bolometric experiments such as CUPID, AMoRE, and TIN.TIN, offering low resolution, scalability, and strong background rejection, are poised to play leading roles. With global coordination and advances in detector sensitivity, the coming decade may bring either the first discovery of $0\nu\beta\beta$ decay or the most stringent constraints yet on lepton number violation and the absolute neutrino mass.

\section{Conclusions and Outlook\label{sec:summary}}

In recent years, advances in cryogenic technologies have led to world-leading sensitivities and first-of-their-kind observations, positioning them at the forefront of rare-event physics. The upcoming generation of experiments, with larger target masses, improved energy resolution, and enhanced background control, is expected to significantly extend discovery reach. Together, these efforts will not only deepen our understanding of fundamental neutrino and dark sector properties, but also probe physics well beyond the Standard Model. A concise overview of the current status, key limitations, and future goals is provided in Table~\ref{tab:future_directions}. With continued innovation and international collaboration, cryogenic detectors are poised to remain a leading platform in the search for new physics in the coming decades.

{\renewcommand{\arraystretch}{1.3}
\setlength{\tabcolsep}{4pt}
\begin{longtable}{p{3 cm}| p{9.0 cm}}
\caption{\label{tab:future_directions}Summary of present status, bottlenecks, and future directions for cryogenic rare-event searches.}\\
\hline
\multicolumn{2}{c}{\rule[12pt]{0pt}{0pt} \textbf{Cryogenic detector technologies}} \\
\hline
\textbf{Key achievements} & Sub-100~eV thresholds demonstrated; dual phonon–ionization and phonon–light readouts established; near tonne-scale bolometric arrays (CUORE) operational. \\
\hline
\begin{minipage}[t]{\linewidth}{\bf Technical\\ bottlenecks}\end{minipage} & Scalability of mK cryogenics; Precise low-energy calibration; limited multiplexed readout; surface-event backgrounds. \\
\hline
\textbf{Next-generation Goals} & Hybrid TES / NTD architectures; superconducting MKIDs / SNSPDs; quantum-limited readout; large-area phonon sensors with single-eV resolution. \\
\hline
\multicolumn{2}{c}{\textbf{WIMP dark matter}}\\
\hline
\textbf{Key achievements} & World-leading sub-GeV limits from CRESST-III and SuperCDMS; low-background operation demonstrated. \\
\hline
\begin{minipage}[t]{\linewidth}{\bf Technical\\ bottlenecks}\end{minipage} & Low energy quenching-factor uncertainties; near-threshold calibration; CE$\nu$NS “neutrino fog”; limited target mass. \\
\hline
\textbf{Next-generation Goals} & SuperCDMS SNOLAB and CRESST-III Phase~2; inelastic (Migdal) and directional searches; multi-target low-mass program. \\
\hline
\multicolumn{2}{c}{\textbf{Axion-like particles \& dark photons}}\\
\hline
\textbf{Key achievements} &  First dedicated cryogenic absorption searches; competitive constraints from SuperCDMS HVeV and EDELWEISS.\\
\hline
\begin{minipage}[t]{\linewidth}{\bf Technical\\ bottlenecks}\end{minipage} & Energy-scale and resolution systematics at eV energies; uncertainties in photoelectric and dielectric response; line-like background features.\\
\hline
\textbf{Next-generation Goals} & High-voltage eV-scale phonon amplification; improved optical modelling; gram- to 100~g-scale detectors with sub-eV threshold reach. \\
\hline
\multicolumn{2}{c}{\textbf{Fractionally Charged Particles (FCPs)}}\\
\hline
\textbf{Key achievements} & Upper limits on intensity across a wide range of $\beta\gamma$ and $q/e$ from CDMSlite, Majorana, CUORE, etc.\\
\hline
\begin{minipage}[t]{\linewidth}{\bf Technical\\ bottlenecks}\end{minipage}  & Low detection efficiency for smaller $q/e$; small acceptance and exposure; background modelling at low ionization.
\\
\hline
\textbf{Next-generation Goals} & Increased detector mass and segmentation; dual heat–light readout for PID; CUPID and SuperCDMS upgrades targeting $f<10^{-8}e$. \\
\hline
\multicolumn{2}{c}{\textbf{Coherent Elastic $\nu$–Nucleus Scattering (CE$\nu$NS)}}\\
\hline
\textbf{Key achievements} & First reactor-based CE$\nu$NS evidence (CONUS+); sub-200~eV$_\mathrm{ee}$ thresholds achieved in Ge detectors.\\
\hline
\begin{minipage}[t]{\linewidth}{\bf Technical\\ bottlenecks}\end{minipage} & Uncertainties in Ge quenching factor; reactor $\bar\nu_e$ flux models; low-energy calibration and background subtraction. \\
\hline
\textbf{Next-generation Goals} & Multi-target calorimeters (NUCLEUS, Ricochet); improved flux characterisation; combined reactor–accelerator measurements to probe BSM couplings. \\
\hline
\multicolumn{2}{c}{\textbf{Neutrinoless Double-Beta Decay (0$\nu\beta\beta$)}}\\
\hline
\textbf{Key achievements} & Tonne-scale bolometric operation demonstrated by CUORE with BI at the $\sim10^{-2}$~counts/(keV$\cdot$kg$\cdot$yr) level and sensitivities up to $3.3\times10^{25}$~yr. CUPID demonstrators (CUPID-0, CUPID-Mo) further achieved BI in the $\sim10^{-3}$~counts/(keV$\cdot$kg$\cdot$yr) range, validating the low-background potential of scintillating bolometers.\\
\hline
\begin{minipage}[t]{\linewidth}{\bf Technical\\ bottlenecks}\end{minipage} & Surface $\alpha$ contamination; cosmogenic activation; isotopic enrichment and mass scaling; calibration at $Q_{\beta\beta}$. \\
\hline
\textbf{Next-generation Goals} & Scintillating bolometers with particle ID (CUPID, AMoRE), targeting BI below $10^{-4}$~counts/(keV$\cdot$kg$\cdot$yr) level; $T_{1/2}^{0\nu}>10^{27}$~yr sensitivity goal; cross-validation with LEGEND and nEXO.\\
\hline
\end{longtable}}

\section{Acknowledgements}
We thank the Department of Atomic energy (DAE), India and the Department of Science and Technology (DST), India for financial support. The work is also funded through the J. C. Bose fellowship of Anusandhan National Research Foundation (ANRF) awarded to Prof. Bedangadas Mohanty. We would like to acknowledge the use of the Garuda and Kanaad HPC cluster facilities at SPS, NISER. We also like to thank Prof. Rupak Mahapatra and his group at Texas A\&M for several collaborative work with us on the cyrogenic detectors. 
 
\bibliographystyle{ws-mpla}
\bibliography{sample}

\end{document}